\documentclass[sn-standardnature]{sn-jnl}

\jyear{2021}%

\theoremstyle{thmstyleone}%
\theoremstyle{thmstyletwo}%

\theoremstyle{thmstylethree}%
\newcommand{\pa}{\partial}

\newcommand{\ga}{\gamma}
\newcommand{\la}{\lambda}
\DeclareUnicodeCharacter{2212}{-}

\begin{document}

\title[Multicriticality]{Multi-criticality and related bifurcation in accretion discs around non-rotating black holes -- an analytical study}


\author[1]{\fnm{Arpan} \sur{Krishna Mitra}}\email{arpankrishnamitra@hri.res.in}
\equalcont{These authors contributed equally to this work.}

\author[1,2]{\fnm{Aishee} \sur{Chakraborty}}\email{aisheechakraborty@hri.res.in}
\equalcont{These authors contributed equally to this work.}

\author[3]{\fnm{Pratik} \sur{Tarafdar}}\email{pratikt@imsc.res.in}
\equalcont{These authors contributed equally to this work.}

\author*[1,2]{\fnm{Tapas} \sur{Kumar Das}}\email{tapas@hri.res.in}

\affil*[1]{\orgdiv{Department of Physics}, \orgname{Harish-Chandra Research Institute}, \orgaddress{\street{Chhatnag Road, Jhunsi}, \city{Allahabad (Prayagraj)}, \postcode{211019}, \state{Uttar Pradesh}, \country{India}}}

\affil[2]{\orgname{Homi Bhabha National Institute}, \orgaddress{\street{Training School Complex, Anushakti Nagar}, \city{Mumbai}, \postcode{400094}, \state{Maharashtra}, \country{India}}}

\affil[3]{\orgdiv{Department of Physics}, \orgname{Institute of Mathematical Sciences}, \orgaddress{\street{IV Cross Street, CIT Campus}, \city{Taramani, Chennai}, \postcode{600113}, \state{Tamil Nadu}, \country{India}}}


\abstract{Low angular momentum, general relativistic, axially symmetric accretion of hydrodynamic fluid onto Schwarzschild black holes may undergo more than one critical transition.
To obtain the stationary integral solutions corresponding to such multi-critical accretion flow, one needs to employ numerical solutions of the corresponding fluid dynamics equations. In
the present work, we develop a {\it{completely analytical}} solution scheme which may be used to find several trans-critical flow behaviours of aforementioned accretion, without explicitly
solving the flow equations numerically. We study all possible geometric configurations of the flow profile, governed by all possible thermodynamic equations of state. We use Sturm's
chain algorithm to find out how many physically acceptable critical points the accretion flow can have, and discuss the transition from the mono to the multi-critical flow profile, and related
bifurcation phenomena. We thus illustrate, completely analytically, the application of certain aspects of the dynamical systems theory in the field of large scale astrophysical flow under
the influence of strong gravity. Our work may possibly be  generalized to  calculate  the maximal number of equilibrium points certain autonomous dynamical systems can have in general.}




\maketitle

\section{Introduction}\label{intro}

Weakly rotating, general relativistic, axially symmetric accretion of hydrodynamic matter onto astrophysical black holes may exhibit multi-transonic behaviour. Subsonic matter starting from infinity may become supersonic after crossing a sonic point located far away from the horizon. Such supersonic matter may encounter a stationary shock and becomes subsonic again. Since the inner boundary condition imposed at the horizon requires the accreting matter to plunge through the horizon supersonically, shock induced subsonic matter becomes supersonic again after crossing another sonic point located relatively close to the black hole's event horizon. One thus obtains two subsonic to supersonic smooth transitions at two respective sonic points, and a supersonic to subsonic discontinuous transition at the shock. Overall profile of such multi-transonic flow may be realized  by studying the stationary integral accretion solutions on the Mach number vs radial distance (measured from the horizon) plot. One needs to take recourse to numerical solution of the corresponding flow equations, i.e., the general relativistic Euler and the continuity equation, to  obtain the stationary integral multi-transonic flow solutions. A large body of work is available in the literature which studies the aforementioned flow profile, either using semi-analytic methods for stationary accretion flow, or by employing complete numerical simulations of full space-time dependent flow structure \cite{fukue83pasj, lu85aa, lu86grg, fukue87pasj, ky94mnras, nakayama94mnras, yk95aa, chakrabarti96mnras, pariev96mnras, lyyy97aa, pa97mnras, ct98apj, cd01mnras, trft02apj, bdw04apj, ny08apj, ny09apj, dc12mnras, bcdn14cqg, dnhbmcbwkn15na, td15ijmpd, abd15grg, sj15mnras, lwwbp16apj, ssnrd16na, scj17mnras, sfd17cqg, bbd17na, shaikh18cqg, td18ijmpd, dsd18na, ddmc18prd, mnd18mnras, sd18jaa, sd18prd, smnd19na, ddmn19mnras, ddn19mnras, pjs19mnras, pjs19hepr, pjc20apj, dmcd20mnras, soa21raa, tmd21prd}. It has usually been observed that there exists, as of now, no general mathematical prescription which may allow one to have a complete analytical understanding of the nature of the overall behaviour of the multi-transonic shocked accretion flow.

Borrowing the methodology from the dynamical systems theory, it has been demonstrated that a physical stationary transonic accretion solution can formally be realized as a critical solution on the phase portrait \cite{rb02pre, ray03mnrasa, ray03mnrasb, rb05arxiv, rb05apj, rb06ijp, rb07cqg, br07apj, bbdr09mnras}. From that point of view, multi-transonic flow solutions can actually be understood as the specific subset of the critical flows for which one obtains multiple critical points. It can be shown that for multi-critical black hole accretion, stationary solution passes through two saddle type and flows around a centre type critical point. Such critical points may either be isomorphic to the sonic points or, for certain special geometric configurations of the flow, may not coincide with the sonic points. In the latter cases, the sonic points can either be numerically obtained by integrating the flow from the corresponding critical points, or an effective sound speed may be introduced which would wean the sonic transition off its distinction from the critical transition.

The correspondence between the phase portraits of the multi-critical solutions of autonomous dynamical systems and the multi-transonic solutions of hydrodynamic flows initiated the effort of studying the multi-transonic properties of accreting black hole systems within a complete analytical framework. It has recently been shown \cite{addn12grg} that using the theory of algebraic polynomials, one can study the transitions from mono-transonic accretion to multi-transonic accretion by computing how many critical points an accretion flow can have for a certain set of initial boundary conditions. Such work studied relativistic accretion flow onto a non-rotating black hole for a particular geometric configuration of axisymmetric accretion governed by polytropic equations of state.

Dynamical properties of the axially symmetric hydrodynamic accretion onto astrophysical black holes are usually studied for three particular geometric configurations \cite{cd01mnras, bcdn14cqg, td18na, tbnd19prd}, namely, flows with constant thickness, the quasi-spherical flow (flows with constant ratio of thickness to radial distance), and flows in hydrostatic equilibrium along the transverse (vertical) direction. For accretion in hydrostatic equilibrium along the transverse direction, the flow thickness is found to be a nonlinear explicit or implicit function of the radial distance. For general relativistic black hole accretion discs, different functional forms of such thickness (disc height) are available in the literature. Details of such configurations have been briefly discussed in consecutive sections. It is also to be noted that apart from a polytropic flow, thermodynamic states of accreting matter are also described by the isothermal equation of state, which provides considerably different flow profile in comparison to the polytropic accretion.

Motivated by the aforementioned diversity in the geometric configurations of the disc structure as well as its thermodynamic properties, we would like to generalize the work presented in \cite{addn12grg} for {\it all} available variants of the accretion disc models governed by both of the thermodynamic equation of states, i.e., polytropic as well as isothermal, respectively. We believe that such a comprehensive treatment will be of greater help to understand certain characteristic features of multi-transonic black hole solutions using fully analytic framework -- which, as we believe, is the main achievement of our paper, that we have been able to provide a comprehensive understanding of the dynamics of weakly rotating sub-Keplerian transonic relativistic flow using the full elegance of analytical treatment.

To accomplish such a task, for certain type of spacetime metric (as elaborated in the next section), using the appropriate co-ordinate systems, we first provide the expression for the energy momentum tensor of inviscid ideal fluid. For low angular momentum sub-Keplerian flow, the effective viscous time scale is considerably larger than the radial infall time-scale, hence inviscid flow is a reasonably valid approximation as we believe. In fact, such effectively inviscid flows can be observed at our own Galactic centre, for detached black hole binary systems fed by accretion from OB stellar wind, or in semi-detached low mass non-magnetic binary systems \cite{is75aa,ln84aa,illarionov88sa,bbckm98mnras,ho99,ia99mnras}.

Vanishing of the four divergence of the energy momentum tensor leads to the derivation of the general relativistic Euler equation and the equation of continuity. Integration of the continuity equation provides the expression of the mass accretion rate which is one of the two constants of motion considered for our system. The other constant of motion is obtained by integrating the Euler equation. One needs to specify the geometrical configuration of the flow in order to compute the mass accretion rate. Mass accretion rates are thus different for different flow geometries. On the other hand, integration of the Euler equation requires the knowledge of the thermodynamic equation of state governing the flow. The constant of motion thus obtained will therefore be different for polytropic and isothermal accretion. While the constant of motion obtained by integrating the polytropic Euler equation turns out to be the total conserved specific energy of the system, such constant obtained for the isothermal flow cannot be directly identified with any exact physical entity. However, even for isothermal flows, the derived conserved quantity has been found to affect the system energy and is used to quantify the amount of energy liberated at isothermal shocks.

For each geometric configuration of the flow, one thus obtains two sets of solutions of the Euler and the continuity equations -- one for flow with such geometry and governed by the polytropic equation of state, and the other for flow with such geometry but governed by the isothermal equation of state. Integral solutions of the Euler and the continuity equations provide two conserved quantities for our flow models. We have three different flow geometries, and out of those three geometries, the flow in hydrostatic equilibrium along the transverse direction has three different expressions for the (radial) distance dependent disc thickness, and all such models can be considered as both polytropic as well as isothermal flows. We thus have five different cases (three different flow geometries with three different disc heights for one particular kind of geometry, i,e., for flow in hydrostatic equilibrium along the transverse direction) of flow structures, each governed either by the polytropic or by the isothermal equation of state. Hence we have ten distinct flow profiles (including the geometric as well as the thermodynamic variants). Out of the aforementioned ten flow models, adiabatic flow in hydrostatic equilibrium along the vertical direction, where the radial distance dependent disc height has been provided by Abramowicz-Lanza-Percival model \cite{ablp}, has been addressed in \cite{addn12grg}, hence we omit the analysis of such model, and we focus on the rest of the models (nine  altogether)  in our present work. 

Once the constants of motions are obtained for a particular model, we derive, by differentiating those first integrals of motion, the equations describing the spatial gradient of the dynamical velocity of the accreting matter as well as that of the sound speed. It is observed that such equations have a one-to-one correspondence with the set of equations that describes a first order autonomous dynamical system. Fixed point analysis is then performed to find out the critical point conditions. Using those critical point conditions, we demonstrate that the conserved quantity obtained by integrating the stationary Euler equation can be expressed as a $n^{\rm th}$ degree polynomial of the critical points $r_c$ with constant co-efficients. The value of $n$ varies for different flow models with different equations of state used to describe the flow. The constant co-efficients of the polynomials turn out to be the functions of the system parameters specifying the flow -- viz. the conserved specific energy ${\cal E}$, specific angular momentum $\lambda$ (which is assumed to be constant for the inviscid flow we consider here) and the polytropic index $\gamma$ (ratio of the specific heat at constant pressure to the specific heat at constant volume, i.e., $\gamma=c_p/c_v$) for the polytropic flow; and the constant specific angular momentum $\lambda$ and the fixed flow temperature (the effective proton temperature) $T$ for the isothermal flow.

Once the explicit form of such a polynomial is obtained for a particular flow model, we use the Sturm's theorem, which is basically a corollary of the Sylvester's theorem \cite{bcr91}, to construct the Sturm's chain algorithm (refer Appendix \ref{secA1}). For certain specific set of $\left[{\cal E},\lambda,\gamma\right]$ or $\left[T,\lambda\right]$, we use the 
algorithm to find out the number of such real roots of the polynomial that are located outside the event horizon of the black hole. Such roots are basically the critical points which in turn are correlated to the sonic points. Hence, although we need to adopt numerical methods to calculate the exact location of the sonic point(s) for accretion flow, we can predict whether the flow will be monotransonic or multi-transonic just by computing the number of roots completely analytically. In this process, we can also realize how the continuous and gradual change of the initial boundary condition (set of flow parameter values) leads to the mono to multi-transonic transition (or vice versa) of the accretion flow. Such transition gets manifested as bifurcation phenomena on the parameter space specified by $\left[{\cal E},\lambda,\gamma\right]$ or $\left[T,\lambda\right]$. The bifurcation curves on the parameter space correspond to the transition boundaries of the regions across which the number of critical points changes from one (mono) to three (multi) or vice-versa. The multi-transonic behaviour for {\it all} kind of flow geometries governed by all thermodynamic equations of state can thus be realized in our work within completely analytical framework in a comprehensive way.

\section{Background space-time geometry and the first integrals of motion}
We consider the accretion to occur in a (3+1) stationary axially symmetric space-time having two killing vectors commuting with each other. The local time-like  killing vector $\xi^{\mu} \equiv (\frac{\pa}{\pa t})^{\mu}$ plays the role of generator of the stationary behaviour, while $\phi^{\mu}\equiv (\frac{\pa}{\pa \phi})^{\mu}$ generates the axial symmetry. We have done all the measurements in natural units(G=c=1). $M_{\rm{bh}}$, the black hole mass, is also scaled to $1$ for notational simplicity and can be substituted back dimensionally.
We consider the energy momentum tensor  of an ideal fluid to be $$T^{\mu\nu}=(\epsilon + p)u^{\mu}u^{\nu}+ p g^{\mu\nu},$$  defined in the aforementioned spacetime. Such a spacetime is described by cylindrical Boyer-Lindquist coordinate in the equatorial plane $(z =0)$ as defined below:
\begin{equation}
    \label{met}
    ds^2= g_{\mu\nu}dx^{\mu}dx^{\nu}=-\frac{r^2 \Delta}{A}dt^2 +\frac{A}{r^2}(d\phi -\omega dt)^2 +\frac{r^2}{\Delta}dr^2 +dz^2
\end{equation}

where $$ \Delta= r^2 -2r + a^2, ~~~~ A= r^4 +r^2a^2 +2r a^2, ~~~~ \omega=\frac{2ra}{A},$$ \\$a$ being the Kerr parameter.
\noindent
 When we consider accretion onto non-rotating black hole, we substitute $a =0$. With this condition, we have, $$\Delta=r(r-2),~~A=r^4, ~~\text{and} ~~\omega =0.$$ The metric components are then given as,
 \begin{equation}
     \label{meel}
     g_{rr}=\frac{r}{(r-2)},~~~~g_{tt}=-\frac{(r-2)}{r},~~~~g_{\phi \phi}=r^2,~~~~g_{t\phi}=g_{\phi t}=0.
 \end{equation}
 
 \subsection{First integrals of motion}
 We have two types of accretion flow governed by two different thermodynamic equations of states, namely, the polytropic, in which the energy conservation holds and the isothermal, i.e. a flow with constant temperature.  
 The relativistic Euler equation for polytropic accretion is constructed from the conservation equation of the energy momentum tensor,
 \begin{equation}
     \label{coneu}
     T^{\mu\nu}_{;\nu}=0.
 \end{equation}
 On top of that we have another conservation equation, namely the continuity equation
 \begin{equation}
     \label{conti}
     (\rho u^{\mu})_{;\mu}=0
 \end{equation}
 in fluid context.
 \subsubsection{Integral solution of the linear momentum conservation equation}
 \subsubsection*{Polytropic accretion}
 Contraction of eqn. \eqref{coneu} with the specified Killing vectors $\phi^{\mu}$ and $\xi^{\mu}$ produces angular momentum per baryon $h u_{\phi}$ and relativistic Bernoulli's constant $h u_{t}$ ($h$ is the specific enthalpy of the system) respectively, which can be shown to be conserved. 
 $hu_{t}$ can be identified with $\mathcal{E}$, the total specific energy of the ideal general relativistic fluid scaled in units of the rest-mass energy. \\
\noindent
We define the specific angular momentum ($\lambda$) and the angular velocity ($\Omega$) as,
\begin{equation}
    \label{angv}
    \lambda = -\frac{u_{\phi}}{u_{t}}
\end{equation}
 \begin{equation}
     \label{angm}
     \Omega = \frac{u^{\phi}}{u^{t}}= -\frac{g_{t\phi} + \lambda g_{tt}}{g_{\phi\phi}+\lambda g_{t\phi}}
 \end{equation}
 Using the normalization condition $u^{\mu}u_{\mu}=-1$ we obtain,
 \begin{equation}
     \label{ut}
     u_{t}=\sqrt{\frac{g_{t\phi}^{2}-g_{tt}g_{\phi\phi}}{(1-\lambda\Omega)(1-u^2)(g_{\phi\phi}+\lambda g_{t\phi})}}
 \end{equation}
 The corresponding expression for the conserved specific energy $\mathcal{E}$ is therefore given by,
 \begin{equation}
     \label{eneut}
     \mathcal{E}=\frac{\gamma -1}{\gamma -(1+ c_{s}^{2})}\sqrt{\frac{g_{t\phi}^{2}-g_{tt}g_{\phi\phi}}{(1-\lambda\Omega)(1-u^2)(g_{\phi\phi}+\lambda g_{t\phi})}},
 \end{equation}
 where $\gamma$ is the ratio of the specific heats at constant pressure and constant volume $\left(\frac{C_{p}}{C_{v}}\right)$ and $c_{s}$ is the local sonic velocity.
\noindent
 It is evident that the specific energy does not have explicit dependence on the geometrical configuration of the accretion flow. It depends on the spacetime geometry. 
 \noindent\\
 \subsubsection*{Isothermal accretion}
 In case of isothermal accretion, energy dissipation occurs to maintain a constant temperature. Hence, the total energy cannot be conserved. However we may derive a  conserved quantity upon integration of the relativistic Euler equation, which is different from the total energy of the system. 
 
\noindent 
 The isotropic pressure is directly proportional to the specific energy of the fluid 
 \begin{equation}
     \label{pree}
     p=c_{s}^{2}\epsilon
 \end{equation}
 \noindent
 The energy-momentum conservation equation we obtain by setting the 4-divergence (covariant derivative w.r.t. $\nu$) of \eqref{coneu} to be zero is,

\begin{equation}
\label{euis}
p_{,\nu}(g^{\mu\nu}+u^{\mu}u^{\nu})+(p+\epsilon) u^{\nu}u^{\mu}_{;\nu}=0
\end{equation}

Using \eqref{pree} we get the general relativistic Euler equation for isothermal accretion.
\begin{equation}
\label{euio}
\frac{c_{s}^2}{\rho}\rho_{,\nu}(g^{\mu\nu}+u^{\mu}u^{\nu})+u^{\nu}u^{\mu}_{;\nu}=0
\end{equation}
We introduce the irrotationality condition here, $\omega_{\mu\nu}=0~\text{with,}~ \omega_{\mu\nu}=l^{\alpha}_{\mu}l^{\beta}_{\nu}v_{[\alpha;\beta]}$. Here, $l^{\alpha}_{\mu}=\delta^{\alpha}_{\mu}+u^{\alpha}u_{\mu}$ is the projection operator acting along the normal direction of $u^{\mu}$, and $u_{[\alpha;\beta]}=\frac{1}{2}(u_{\beta;\alpha}-u_{\alpha;\beta}).$
We substitute this condition into \eqref{euio} to obtain
\begin{equation}
\label{iscon}
\pa_{\nu}(u_{\mu}\rho^{c_{s}^{2}})-\pa_{\mu}(u_{\nu}\rho^{c_{s}^{2}})=0
\end{equation}
The time component of \eqref{iscon} shows, for isothermal irrotational flow, $u_{t}^2\rho^{2c_{s}^{2}}= \xi$ is a conserved quantity.
Hence, $\xi$ is the first integral of motion isothermal accretion, which cannot be identified with the specific energy of the system.

\subsubsection{Integral solution of the mass conservation equation}
The continuity equation \eqref{conti} gives us,
\begin{equation}
    \label{intc}
    \frac{1}{\sqrt{-g}}(\sqrt{-g}\rho u^{\mu})_{,\mu}=0,
\end{equation}
where $g \equiv \rm{det}(g_{\mu\nu})$. We now have,
\begin{equation}
    \label{conin}
    d^4 x (\sqrt{-g}\rho u^{\mu})_{,\mu} =0
\end{equation}
with $\sqrt{-g}d^4 x$ being the covariant volume element. The $z$ component of the velocity $u^{z}$ (in cylindrical coordinates) or the polar component $u^{\theta}$ (in spherical polar coordinates) are assumed to be negligible w.r.t the transformed radial component $u^{r}$. With this assumption we have,
\begin{equation}
    \label{ycont}
    \pa_{r}(\sqrt{-g}\rho u^{r})dr d\theta d\phi =0;~~~~~ \pa_{r}(\sqrt{-g}\rho u^{r})dr dz d\phi =0
\end{equation}
\noindent
for stationary flow in spherical and cylindrical polar coordinates, respectively.
\noindent\\
The equation for spherical polar coordinates is integrated for $\phi$ from $0$ to $2\pi$ and $\theta$ from $-H_{\theta}$ to $H_{\theta}$, $\pm H_{\theta}$ being the  corresponding values of the coordinates above and below the equatorial plane, respectively, for the local half thickness $H,$ to obtain the conserved quantity $\dot{M}$, the mass accretion rate, when $\theta =\frac{\pi}{2}$(i.e. we consider the value on the equatorial plane).
\noindent
The mass accretion rate $\dot{M}$ depends on the matter geometry configurations. We here present a general expression for it,
\begin{equation}
    \label{Mg}
    \dot{M}=\rho u^{r} \mathcal{A}(r),
\end{equation}
where $\mathcal{A}(r)$ stands for the $2D$ surface area through which the steady state inbound mass flux is calculated.

\section{Disc geometries}
We here follow the footsteps of the existing literature to assume that the axially symmetric accretion flow has a well defined local thickness, and the central plane of the flow coincides with the equatorial plane of the black hole. Below we briefly discuss various geometric configurations of the axisymmetric flow.

\noindent
\subsection{Flow with a constant height:}

\noindent
For constant height (CH hereafter) flows, height or half-thickness $(H)$ of accretion disc is constant throughout. $H$ does not depend on the radial distance, i.e.

\begin{equation}
\label{ah}
H(r)= \rm{constant}
\end{equation}

\subsection{Quasi-spherical flow:}

In case of quasi-spherical (qu hereafter) accretion, the flow geometry obeys the conical structure resembling a sphere with cones scooped out from opposite poles. In such geometry, the disc height (H) varies linearly with radial distance (r), i.e.

\begin{equation}
\label{quh}
    H = r \Lambda _{a},
\end{equation}
where $\Lambda_{a}$ is the solid angle subtended by the conical flow at the event horizon which is constant throughout the flow.

\subsection{Flow with hydrostatic equilibrium in vertical direction:}

\noindent
In this model, it is customary to vertically average out physical variables associated with flow (such as pressure p, density $\rho$ etc.) to define such quantities on the equatorial plane \cite{katsu}. The disc height can be obtained by balancing the gravitational force with the pressure gradient. This model can be classified into three subdivisions: 

\subsubsection{Novikov--Thorne model:}

\noindent
The disc height or the half-thickness $H(r)$  for the Novikov-Thorne model \cite{noth} (NT hereafter) is given by, 

\begin{equation}
\label{nothh}
H(r)=\left(\frac{p}{\rho}\right)^{\frac{1}{2}} r^{\frac{3}{2}} \left(1-\frac{3}{r}\right)\left(1-\frac{2}{r}\right)^{-\frac{1}{2}}  
\end{equation} \\

\noindent
\subsubsection{Riffert--Herold model:}

\noindent
For the Riffert-Herold model \cite{rih} (RH hereafter), the half thickness or height of the disc $H(r)$ is given by, 
\begin{equation}
    \label{rhh}
    H(r)=2 \left(\frac{p}{\rho}\right)^{\frac{1}{2}} r^{\frac{3}{2}} \left(1-\frac{3}{r}\right)^\frac{1}{2}
\end{equation}

\noindent
\subsubsection {Abramowicz-Lanza-Percival model:}

\noindent
The  height or half-thickness of the disc $H(r)$ for the Abramowicz-Lanza-Percival model \cite{ablp} (ALP hereafter) is given by:

\begin{equation}
\label{abrh}
H(r)= \frac{r^{2}c_{s}}{\lambda}\left(\frac{2(1-u^{2})(1-\frac{\lambda^{2}}{r^{2}}(1-\frac{2}{r}))(\gamma-1)}{\gamma(1-\frac{2}{r})(\gamma-(1+c_{s}^{2}))}\right)^\frac{1}{2}
\end{equation} \\

\noindent

\section{Polytropic accretion}
The equation of state governing a polytropic fluid is given by,
\begin{equation}
\label{pol}
p=K\rho^{\gamma},
\end{equation}
where $p$ is the pressure of the accreting fluid, $\ga$ is the polytropic index which is assumed to be constant throughout the accretion process and $K$ gives us the measure of specific entropy, provided, no additional generation of entropy occurs in the process.
\noindent
Specific enthalpy can be formulated as,
\begin{equation}
\label{spa}
h= \frac{\epsilon +p}{\rho}
\end{equation}
with, $\epsilon$ being the internal energy density of the system measured by the relation
\begin{equation}
\label{epa}
\epsilon = \rho +\frac{p}{\gamma -1}
\end{equation}
The speed of acoustic propagation in a relativistic polytropic flow is obtained as,
\begin{equation}
\label{soa}
c_{s}^2=\frac{\pa p}{\pa \epsilon}\bigg\vert_{h}
\end{equation}
\noindent
Simple manipulation of \eqref{epa} and \eqref{soa} gives,
\begin{equation}
\label{pode}
\rho =\left[\frac{c_{s}^{2}(\gamma -1)}{\gamma K(\gamma -1 -c_{s}^{2})}\right]^{\frac{1}{\gamma -1}}
\end{equation}

\noindent
In polytropic accretion, we have two first integrals of motion, namely, the conserved specific energy $\mathcal{E}$, and the mass accretion rate $\dot{M}$, which are given by,
\begin{equation}
\label{enea}
\mathcal{E}= \frac{(\ga -1)r}{\ga -1 - c_{s}^{2}}\sqrt{\frac{(r-2)}{(1- u^2)(r^3-\la^2(r-2))}}, \text{ and}
\end{equation}

\begin{equation}
\label{masa}
\dot{M}=\rho v^r {\it{A}}(r).
\end{equation}
\noindent
This energy expression (obtained by integrating the stationary part of the Euler equation) holds good for all the accretion disc geometries. On the other hand, the mass accretion rate obtained by integrating the stationary part of the continuity equation is explicitly dependent on the specific disc geometry. 
\noindent\\
We are going to calculate the radial derivative of the dynamical velocity $u$ for different disc geometries. For a physically realisable transonic flow, $\frac{du}{dr}$ has to be smooth at every point within the astrophysical domain. In other words, if spatial derivative of dynamical velocity becomes undefined, i.e. if the denominator $(\mathcal{D})$ equals to zero, the numerator($\mathcal{N}$) should also be zero in order to keep smoothness of the flow intact. Now, at the critical points, $\frac{du}{dr}$ can assume multiple values depending on the specific set of system parameters. Mathematically, this would be equivalent to the denominator of $\frac{du}{dr}$ going to zero. However, for the flow to remain smooth in addition, as justified above, the numerator should also be equated to zero at the critical points. Or in other words, the criticality condition demands that both the numerator as well as denominator of the spatial gradient of the advective flow velocity be equal to zero. This condition when fed into \eqref{enea} gives us the required polynomial equation whose solutions are the critical points ($r_c$). 
\noindent\\
These polynomials, as will be shown in the subsequent sections, are usually (except in a few special cases) of order $n>4$ and hence, not analytically solvable. So, in order to obtain the number of real roots of the polynomial equations for a specified range of parameters within relevant astrophysical domain, we need to use the Sturm method as discussed earlier \cite{addn12grg}. 
\noindent\\
The phrase `astrophysical domain' or `astrophysically relevant domain' in the context of flow parameters may demand some explanation. The specific energy $\mathcal{E}$ of the system includes the rest mass energy and it is scaled by the same. Hence, a system with $\mathcal{E} = 1$, would signify, zero thermal energy at an infinitely distant point  which is not a physically acceptable boundary condition and $\mathcal{E} \lt 1$ would mean negative initial energy, in which case, presence of dissipation would be necessary in order to get solutions that include positive energy solutions. But, in our system, we have considered inviscid flow, and hence no such sources of dissipation exist. While $\mathcal{E} \gt 1$ produces all physically acceptable solutions, $\mathcal{E} \gt 2$ represents very high thermal energy initial states, which, in general, are not usual in black hole accretion. The above discussion dictates us to consider the specific energy values lying in the range  $1\lesssim \mathcal{E} \lesssim 2$. Here we are dealing with axisymmetric accretion well within the Keplerian regime. This settles the range of the specific angular momentum as, $ 0\lt \lambda \leq 4$. In case of isothermal accretion, we have, the polytropic index $\gamma = 1$. While $\gamma \lt 1$ is not a physically acceptable value in accretion astrophysics, $\gamma \gt 2$ would imply super-dense plasma with non-negligible magnetic field in addition to a direction dependent anisotropic pressure. Since we are not considering general relativistic magneto-hydrodynamic equations, hence our choice of values for $\gamma$ shall remain fixed below $2$. Again, following the existing black hole accretion literature (Frank et al. [1992]), we can constrain the polytropic index further within ultra-relativistic and non-relativistic bounds as $4/3\leq \gamma \leq 5/3 $ \cite{tbnd19prd}. Thus, the astrophysical limits of our flow parameters are given by, $[1\lesssim \mathcal{E} \lesssim 2, 0\leq \lambda \leq 4, 4/3\leq \gamma \leq 5/3 ]$.

\subsection{Constant height model}
The mass accretion rate for this specific disc geometry (using \eqref{masa}) is given by
 
 \begin{equation}
 \label{poch}
 \dot{M}=2\pi\rho \frac{u \sqrt{1-\frac{2}{r}}}{\sqrt{1-u^{2}}}rH
 \end{equation}
\noindent 
which remains constant throughout the flow.  

\noindent
These two equations \eqref{enea} and \eqref{poch} contain three unknown quantities $\rho , ~ u, \rm{and} ~c_{s}$, which are functions of $r$. Hence, one needs to eliminate one of the three variables by expressing it in
terms of the other two. We here, would like to express $\rho$ in terms of $u~ \rm{and} ~c_{s}$, as we are interested in studying the profile for the
radial Mach number $M_{c}=\frac{u}{c_{s}}$, to get the  location of the sonic points. For this purpose, we make use the  transformation
\begin{equation}
    \label{spen}
    \dot{\mathcal{M}}=\dot{M}(K\gamma)^{\frac{1}{\gamma-1}}
\end{equation}

$\dot{\mathcal{M}}$ may be interpreted as a
measure of the net inbound entropy flux of the fluid and hence, can be defined
as the stationary entropy accretion rate.
This entropy accretion rate was proposed for the first time in \cite{maa, blabla}.
\noindent

Using \eqref{enea}, \eqref{poch} and \eqref{spen}, following the aforementioned method we derive an expression of $\frac{du}{dr}$ which is of the form,

\begin{equation}
 \label{dudr}
\frac{du}{dr}= \frac{c_{s}^{2}(\frac{r-1}{r(r-2)})-f(r, \la)}{\frac{u^2 -c_{s}^{2}}{u(1-u^{2})}}= \frac{\mathcal{N}}{\mathcal{D}}
\end{equation}
 where, 
 
\begin{equation}
    \label{fffffff}
    f(r,\la)=\frac{r^{3}-\la^{2}(r-2)^{2}}{r(r-2)(r^3 - \la^2 (r-2))}
\end{equation}
 
\noindent
Inspecting the denominator $\mathcal{D}$ we can readily conclude that the critical points coincide with the sonic points here, as,

\begin{equation}
\label{con}
\mathcal{D}=0\implies u\mid_{r_{c}} = c_{s}
\end{equation}
\noindent
and, dictated by the argument that $u$ has to be a smooth function of the radial distance, we have $\mathcal{N}=0$ at the critical points. 
 The advective velocity and/or the local sound speed at the critical points can be calculated from this condition.
 \begin{equation}
 \label{ucri}
u_{c}=c_{s}\big\vert_{c}=\sqrt{\frac{f(r, \lambda)}{\frac{r-1}{r(r-2)}}}\bigg\vert_{r_{c}}
 \end{equation}

\noindent
When \eqref{ucri} is fed in \eqref{enea} it produces an $\mathcal{O}(\sim 11)$ polynomial equation in $r_{c}$.\footnote{for the explicit values of the coefficients of the polynomial see Appendix \ref{secA2.1}.}
\begin{equation}
\Sigma_{i=0}^{11} a_{ch_{i}}{r_{c}}^{i}=0
\end{equation}

\begin{figure}
	\begin{center}
		\includegraphics[scale=0.8]{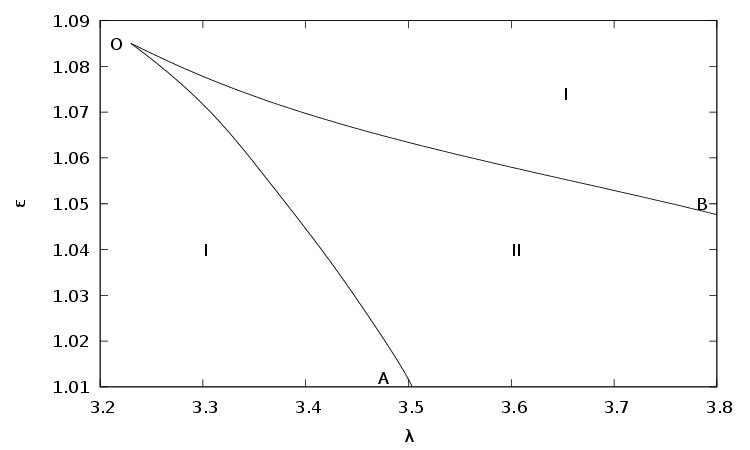}
		\caption{Polytropic accretion in constant height discs and related bifurcation phenomena}
		\label{chpol}
	\end{center}
\end{figure}

Here, the number of critical points is equal to the number of real roots (located outside the horizon) of the above polynomial equation. Now, applying Sturm analysis on the above polynomial equation, we obtain the number of real roots in a certain range of $\lambda$ and $\mathcal{E}$. The corresponding bifurcation diagram to describe the flow qualitatively in $\lambda$ -- $\mathcal{E}$ space has been shown in figure \ref{chpol}. 

\noindent
In figure \ref{chpol}, region I denotes the mono-critical subset of the parameter space where only one critical point exists. Region II denotes the multi-critical subset where three critical points exist. Here, OA and OB curves separate these two regions from one another. The interior of the region OAB (bounded by OA and OB) specifies the multi-critical region II. For example, at $\mathcal{E}= 1.06$, in the range of $\lambda\leq3.34$ (region I) there are mono-critical solutions of $f(r_{c})$ and in the range $3.34\leq\lambda\leq3.56$ there are multi-critical solutions of $f(r_{c})$ (region II) reverting back to mono-critical solutions in the range $\lambda\geq 3.56$ (region I). Basically at first, in region I (upto $\lambda$ = 3.34), there exists only one real root, which means there is a saddle point through which the physical flow occurs. At $\lambda$ = 3.34, that saddle point splits into two different saddle points and in between these two, a center type critical point is created. As a result, we get the number of real roots as 3. This continues up to $\lambda=3.56$ (region II). After $\lambda$ attains a value 3.56, the inner saddle point goes beyond our physically acceptable range of $r_c$ (inside the horizon) and only one critical point through which physical flow can occur (the outer saddle point) remains. So, we get back a mono-critical region (region I). It is to be noted that the centre-type critical point might still be present in the physical range of $r_c$ and technically, region I (to the right of curve OB) should in that case be marked separately as a bi-critical region. However, the centre-type critical point does not allow physical flow trajectory to pass through it. Hence, it is not of our interest and flow with such a phase topology is essentially mono-transonic. This saddle-centre bifurcation occurs for each and every value of $\mathcal{E}$ in the given range of $\mathcal{E}$ in a similar manner. Thus accretion flow continues from a distant region to close proximity of the event horizon and then finally falls onto the black hole.\\

 \subsection{Quasi-spherical flow}
 
\begin{figure}
	\begin{center}
		\includegraphics[scale=0.9]{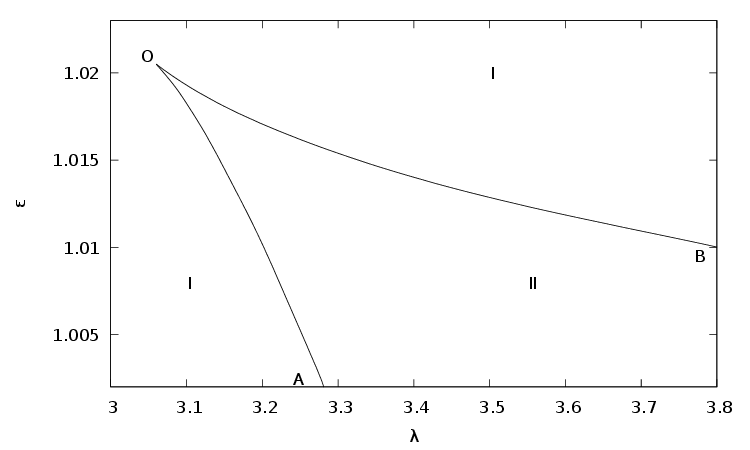}
		\caption{Polytropic accretion in quasi-spherical discs and related bifurcation phenomena}
		\label{qupol}
	\end{center}
\end{figure}

For mass accretion rate in a quasi spherical flow we use \eqref{quh} and \eqref{masa} to obtain,

\begin{equation}
\label{maqu}
 \dot{M}=\Lambda\rho \frac{u \sqrt{1-\frac{2}{r}}}{\sqrt{1-u^{2}}}r^{2}
\end{equation}
with $\Lambda$ being the geometrical solid angle factor.
\noindent
From \eqref{enea},  \eqref{spen} and \eqref{maqu}, and finally using \eqref{fffffff} we arrive at the following expression,

\begin{equation}
\label{poqu}
\frac{du}{dr}= \frac{c_{s}^{2}(\frac{2r-3}{r(r-2)})-f(r, \la)}{\frac{u^2 -c_{s}^{2}}{u(1-u^{2})}}= \frac{\mathcal{N}}{\mathcal{D}}
\end{equation} 
\noindent
Now, as $u$ is a smooth function of $r$,  null value of the denominator at some points in the range implies that the numerator has to be zero simultaneously, which in turn gives us the mathematical condition for critical points. \\
\noindent
Inspecting the denominator $\mathcal{D}$ we can readily conclude that the critical points coincide with the sonic points in this case as well,

\begin{equation}
\label{con1}
\mathcal{D}=0\implies u\mid_{r_{c}} = c_{s}
\end{equation}
\noindent
Again following the same argument as in the previous case, we have $N=0$ at the critical points. 
Hence, the advective velocity and/or the local sound speed at the critical points can be calculated from this condition.
 \begin{equation}
 \label{ucri1}
u_{c}=c_{s}\big\vert_{c}=\sqrt{\frac{f(r, \lambda)}{\frac{2r-3}{r(r-2)}}}\bigg\vert_{r_{c}}
 \end{equation}
\noindent
When \eqref{ucri1} is fed into \eqref{enea} it produces an $\mathcal{O}(\sim 11)$ polynomial equation in $r_{c}$.\footnote{for the explicit expression of the coefficients of the polynomial see Appendix \ref{secA2.2}.}
\begin{equation}
\Sigma_{i=0}^{11} a_{qu_{i}}{r_{c}}^{i}=0
\end{equation}
Here, the number of critical points is equal to the number of real roots (located outside the horizon) of the above polynomial equation.
Now applying Sturm analysis on the above polynomial, we obtain the number of real roots in a certain range of $\lambda$ and $\mathcal{E}$. The bifurcation diagram to describe the flow qualitatively in $\lambda$ -- $\mathcal{E}$ space is shown in figure \ref{qupol}. \\

\noindent
In figure \ref{qupol}, the region I denotes the mono-critical region  and region II denotes the multi-critical region.  Curves OA and OB represent the left and the right boundary (between region I to II).  The interior of the wedge shaped region OAB specifies the multi-critical region II. For example, at $\mathcal{E}= 1.014$, in the range of $\lambda\leq3.16$ (region I) there exists only one physically acceptable solution of $f(r_{c})$ and in the range $3.16\leq\lambda\leq3.40$ there are three such solutions of $f(r_{c})$ (region II) and in the range $\lambda\geq3.40$ there again exists one physically realisable solution of $f(r_{c})$ (region I). 

 \subsection{Flow in vertical hydrostatic equilibrium}

\subsubsection{Novikov-Thorne accretion flow model}

\begin{figure}
	\begin{center}
		\includegraphics[scale=0.9]{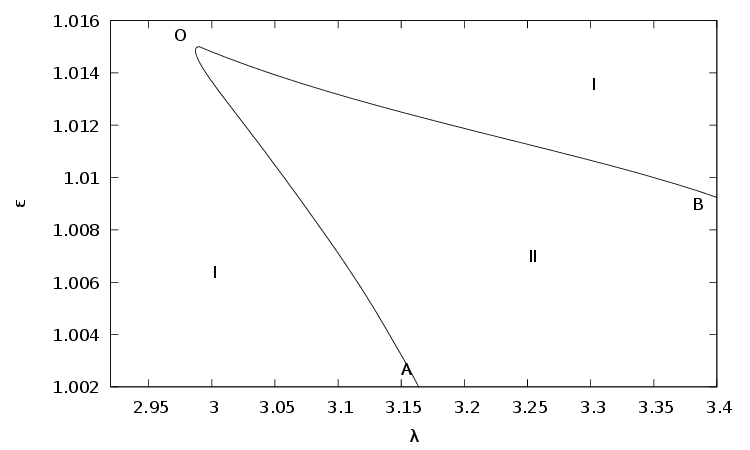}
		\caption{Polytropic accretion in NT flow model and related bifurcation phenomena}
		\label{ntpol}
	\end{center}
\end{figure}

Here the mass accretion rate is given by,

\begin{equation}
\label{mpoly}
\dot{M}=4\pi\rho r \frac{u \sqrt{1-\frac{2}{r}}}{\sqrt{1-u^{2}}} H_{NT}(r).
\end{equation}
\noindent
Using the expression for height of the accretion disc $H_{NT}(r)$, given in \eqref{nothh}
\begin{equation}
\label{mntpo}
\dot{M}=4\pi\rho c_{s} r^{\frac{5}{2}}\frac{u\sqrt{1-\frac{3}{r}}}{\sqrt{1-u^{2}}}
\end{equation}
From \eqref{enea}, \eqref{spen}, \eqref{fffffff} and \eqref{mntpo} we get the following expression for $\frac{du}{dr}$:

\begin{equation}
\label{dent}
\frac{du}{dr}= \frac{c_{s}^{2}(\frac{5r-12}{r(r-3)(\ga +1)})-f(r, \la)}{\frac{u^2(\ga +1) -2c_{s}^{2}}{u(1-u^{2})(\ga + 1)}}= \frac{\mathcal{N}}{\mathcal{D}}
\end{equation}
\noindent
We use the same argument of $u(r)$ being a smooth function of $r$ to obtain the critical point conditions, 
\begin{equation}
\label{crvnt}
u\big\vert_{r_{c}}=\sqrt{\frac{2}{\gamma+1}}c_s\bigg\vert_{r_{c}},
\end{equation}

\begin{equation}
\label{ntn}
c_{s}\big\vert_{r_{c}}=\sqrt{\frac{f(r, \lambda)}{\frac{5r-12}{r(r-3)(\gamma+1)}}}\bigg\vert_{r_{c}}.
\end{equation}
It should be noted that the critical and sonic points do not turn out to be identical in this case. When \eqref{ntn} is fed into \eqref{enea} it produces an $\mathcal{O}(\sim 15)$ polynomial equation in $r_{c}$\footnote{for the explicit expressions of the coefficients of the polynomial see Appendix \ref{secA2.3}.}

\begin{equation}
\Sigma_{i=0}^{15} a_{NT_{i}}x^{i}=0
\end{equation}
\noindent
Here we know that the number of critical points equals to the number of real roots (located outside the horizon) of the  polynomial. Now applying Sturm analysis on the derived polynomial equation, we compute the bifurcation diagram in $\lambda$ - $\mathcal{E}$ space to describe the accretion flow qualitatively. The respective diagram is shown in figure \ref{ntpol}. \\ 

\noindent

In figure \ref{ntpol}, the region I denotes the mono-critical region where only one critical point exists and region II denotes the multi-critical region where three critical points exist. Here, OA and OB curves represent the left and right boundaries respectively (between region II to I). The bounded area OAB  presents the multi-critical region II. As an example, we consider a particular value of specific energy  $\mathcal{E}= 1.012 $, in the range of $\lambda\leq3.02$ (region I) there is a mono-critical solution of $f(r_{c})$ and in the range $3.02\leq\lambda\leq3.19$ there is a multi-critical solution of $f(r_{c})$ (region II) and in the range $\lambda\geq 3.19$ there again exists a mono-critical (mono-transonic, to be precise) solution of $f(r_{c})$ (region I). 

\subsubsection{Riffert-Herold accretion flow model}
 
 Here, mass accretion rate is given by,
\begin{equation}
\label{marh}
\dot{M}=4\pi\rho r \frac{u \sqrt{1-\frac{2}{r}}}{\sqrt{1-u^{2}}} H(r)
\end{equation}
\noindent
Now, substituting the expression of $H(r)$ from \eqref{rhh} into \eqref{marh}, we obtain the final expression of mass accretion rate as,
\begin{equation}
\label{rhm}
\dot{M}=8\pi\rho c_{s}r^{\frac{5}{2}}\frac{u \sqrt{(1-\frac{2}{r})(1-\frac{3}{r})}}{\sqrt{1-u^{2}}}
\end{equation}
\noindent
(which remains constant throughout the flow). \\
\noindent
From \eqref{enea}, \eqref{spen}, \eqref{fffffff} and \eqref{rhm} we obtain the expression for $\frac{du}{dr}$ as,
\begin{equation}
\label{rhdr}
\frac{du}{dr}= \frac{c_{s}^{2}(\frac{5r^{2}-20r+18}{r(r-2)(r-3)(\ga +1)})-f(r, \la)}{\frac{u^2(\ga +1) -2c_{s}^{2}}{u(1-u^{2})(\ga + 1)}}= \frac{\mathcal{N}}{\mathcal{D}}
\end{equation}
In the equation \eqref{rhdr}, equating $\mathcal{D}=0$, we obtain the equation connecting the local advective velocity with the local sonic speed at the critical radius $r_c$.
\begin{equation}
\label{rhcu}
u\big\vert_{r_{c}}=\sqrt{\frac{2}{\ga +1}}c_s \bigg\vert_{r_{c}}
\end{equation}
\noindent
and the other critical point condition $\mathcal{N}=0$, gives us,
\begin{equation}
\label{rhis}
c_{s}\big\vert_{r_{c}}=\sqrt{\frac{f(r, \la)}{\frac{5r^{2}-20r+18}{r(r-2)(r-3)(\ga +1)}}}
\end{equation}
\noindent
Substituting this result in \eqref{enea} generates a polynomial equation of $\mathcal{O}(\sim 14)$ in the critical radius $r_c$\footnote{for the explicit expressions of the coefficients of the polynomial see Appendix \ref{secA2.4}.}.
\begin{equation}
\label{rhpolyy}
\Sigma_{i=0}^{14} a_{RH_{i}}x^{i}=0
\end{equation}

\begin{figure}
	\begin{center}
		\includegraphics[scale=0.9]{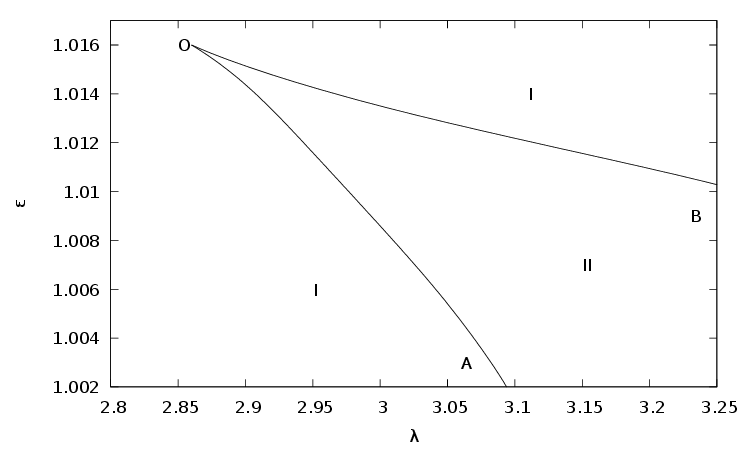}
		\caption{Polytropic accretion in RH flow model and related bifurcation phenomena}
		\label{rhpol}
	\end{center}
\end{figure}

We know that, the number of critical points is equal to the number of real roots (lying outside the horizon) of the above polynomial equation. Applying Sturm analysis on the derived polynomial, we obtain a bifurcation diagram in $\lambda$ - $\mathcal{E}$ space to describe the accretion flow qualitatively. The respective diagram is shown in figure \ref{rhpol}. \\

\noindent
In figure \ref{rhpol}, the region I denotes the mono-critical region where only one critical point exists and region II denotes the multi-critical region where three critical points exist. Here, OA curve represents the left boundary and curve  OB represents the right boundary (between region I to II). Interior of the OAB region (bounded by OA and OB) depicts the multi-critical region II. Now, for example, at $\mathcal{E}=1.012$, in the range of $0\leq\lambda\leq2.94$ (region I) there is a mono-critical solution of $f(r_{c})$ and in the range $2.94\leq\lambda\leq3.10$ there is a multi-critical solution of $f(r_{c})$ (region II) and in the range $\lambda \geq 3.10$ there again exists a mono-critical (one physically realisable critical point) solution of $f(r_{c})$ (region I). The multi-critical range of polytropic flow parameters for different disc geometries are provided in table $1$.

\begin{table}
\begin{center}
\begin{tabular}{|c| c| c| c|}
\hline \hline
Disc geometry  & $\lambda$ &$\mathcal{E}$ & $\gamma$~~ \\ \hline

CH & 3.50-3.80 & $1.01 - 1.085 $ & $\frac{5}{3}$\\ \hline

Quasi-spherical & 3.28-3.80 & $ 1.001-1.020$ & $\frac{5}{3}$\\ \hline

NT & 3.16-3.40 & $1.002 - 1.015$ & $\frac{5}{3}$\\   \hline

RH & 3.09 - 3.25 & $1.002 - 1.016$ & $\frac{5}{3}$\\ \hline
\end{tabular}
\caption{\centering Multi-critical bifurcation region for polytropic accretion in different disc geometries}
\end{center}
\label{tabpol}
\end{table}

\section{Isothermal accretion}

We know that in isothermal flow, temperature $T$ remains constant. Hence, the specific energy can no longer be a conserved quantity due to continuous energy exchange with surroundings. In this case, the conserved quantities are mass accretion rate and another quantity ($\xi$), namely, the quasi-specific energy, which may be derived from integration of the stationary part of the relativistic Euler's equation. The expression for $\xi$ is given by, 

\begin{equation}
\label{comi}
\xi= \frac{r^{2}(r-2)}{(r^{3}-\lambda^{2} (r-2))(1-u^{2})} \rho^{2c_{s}^{2}}.
\end{equation}
\noindent
 Here, \textbf{$\rho$} is the density of the accreting fluid; $r$ is the radial distance of accreting material from the event horizon ; $u$ is the advective velocity which has to be a smooth function of the radial distance in order to get a physically relevant accretion flow  and \textbf{$c_{s}$} is the local sound speed.
\noindent
For isothermal accretion, the equation of state for the accreting fluid is given by,

\begin{equation}
\label{clap}
p=\frac{\kappa_{B} }{\mu m_{H}}\rho T
\end{equation}
\noindent
with $m_{H}$ representing the mass of proton, $\mu$ the mean molecular mass of the ionised hydrogen, $T$ being the temperature of the bulk of the ion and $k_{B}$ signifying the Boltzmann's constant.

\noindent
We know for the isothermal case, $\frac{p}{\rho}=c_{s}^{2}$.  So, 

\begin{equation}
\label{cs}
 c_{s}^{2}=\frac{\kappa_{B}}{\mu m_{H}}T
\end{equation}

\subsection{Constant height model}

 The mass accretion rate for the constant height geometry is:
 
 \begin{equation}
 \label{mch}
 \dot{M}=2\pi\rho \frac{u \sqrt{1-\frac{2}{r}}}{\sqrt{1-u^{2}}}rH
 \end{equation}
 \noindent
which remains constant throughout the flow.
\noindent
From \eqref{comi} and \eqref{mch}, using the previously elaborated methodology, we arrive at the following expression for $\frac{du}{dr}$,

\begin{equation}
\label{dudrch}
\frac{du}{dr}= \frac{u(u^{2}-1)[2r^{3}-2\lambda^{2}(r-2)^{2}+c_{s}^{2}(2r^{3}-2\lambda^{2}r+4\lambda^{2})(1-r)]}{r(2-r)(u^{2}-c_{s}^{2})(2\lambda^{2}r-4\lambda^{2}-2r^{3})}=\frac{\mathcal{N}}{\mathcal{D}}
\end{equation}
\noindent
As $u$ is a continuous function of $r$, if the value of denominator $\mathcal{D}$ is zero at some radial distances within the range of physical interest, then at those points, numerator $\mathcal{N}$ also has to be zero. The above argument gives us the required critical point conditions.
\noindent

\begin{figure}
	\begin{center}
		\includegraphics[scale=0.9]{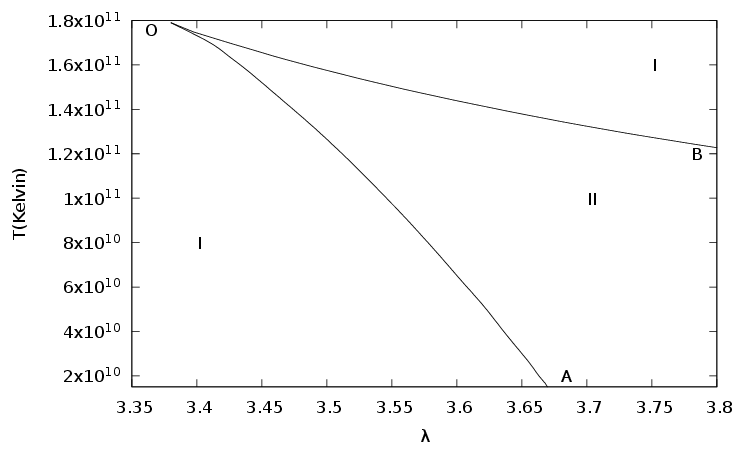}
		\caption{Isothermal accretion in constant hight discs and related bifurcation phenomena}
		\label{chiso}
	\end{center}
\end{figure}

Here, null condition of denominator $\mathcal{D}$ indicates that the critical points shall coincide with the sonic points, i.e.

\begin{equation}
\label{uc}
u\mid_{r_{c}} = c_{s}
\end{equation}
\noindent
Now, since numerator $\mathcal{N}$ also has to be zero at the critical points, so the value of the advective velocity at critical points is given by,

\begin{equation}
\label{crch}
u_{c}=c_{s}\mid_{r_{c}}=\sqrt{\frac{2\lambda^{2}(r-2)^{2}-2r^{3}}{(2r^{3}-2\lambda^{2}r+4\lambda^{2})(1-r)}}
\end{equation}
\noindent
Simplifying the above expression \eqref{crch}, we get the following $4th$-order polynomial equation in $r_{c}$.

\begin{equation}
\label{polych}
f(r_{c})=2c_{s}^{2}r_{c}^{4}-2(1+c_{s}^{2})r_{c}^{3}-2\lambda^{2}(c_{s}^{2}-1)r_{c}^{2}+2\lambda^{2}(3c_{s}^{2}-4)r_{c}-4\lambda^{2}(c_{s}^{2}-2)=0
\end{equation}
\noindent
Therefore for a given set of $T$ - $\lambda$, the corresponding critical points can be obtained by solving for the above polynomial equation\footnote{$c_{s}^2$ can be substituted in terms of $T$ using eqn. \eqref{cs}.}.
 Now applying Sturm analysis on the derived polynomial \eqref{polych} we get the number of real roots (lying outside the horizon) in a certain range of $\lambda$ and $T$ and generate a bifurcation diagram in $\lambda$ - $T$ space to describe the flow qualitatively. The corresponding diagram is shown in figure \ref{chiso}. 

	In figure \ref{chiso}, region I denotes the mono-critical region where only one critical point exists and region II denotes the multi-critical region where three critical points exist. Here, curve OA represents the left boundary (between region I and II) and OB represents the right boundary (between region II and I). Region OAB (bounded by OA and OB) specifies the multi-critical region II. Now, for example, at T = 1.4 $\times 10^{11}$ K, in the range of $0\leq\lambda\leq3.48$ (region I) there exist mono-critical solutions of $f(r_{c})$, while in the range $3.48\leq\lambda\leq3.63$ there are multi-critical solutions of $f(r_{c})$ (region II) reverting back to mono-critical solutions in the range $\lambda\geq3.63$ (region I). As explained in the previous sections, region I (upto $\lambda$ = 3.48), contains only one real root of the given polynomial equation in $r_c$. However, at $\lambda$ = 3.48, two additional critical points are generated. As a result, the number of real roots increases to $3$. This continues upto $\lambda=3.63$ (region II). After $\lambda$ attains a value 3.63, one of the two saddle points (inner), and in some cases the centre-type critical point as well, goes beyond the physically acceptable range of radial distance, leaving behind a single critical point through which physical flow can occur. Thus, we effectively get back a mono-critical region (region I). This saddle--centre bifurcation occurs over a specific range of $T$ (depending on the flow geometry) in a similar manner.

\subsection{Quasi--spherical flow}
 
The mass accretion rate of this accretion flow geometry is given by,

\begin{equation}
\label{mqs}
 \dot{M}=\Lambda_{\text{iso}}\rho \frac{u \sqrt{1-\frac{2}{r}}}{\sqrt{1-u^{2}}}r^{2}
\end{equation}

where $\Lambda_{\text{iso}}$ is the geometrical solid angle factor subtended by the disc at the horizon and it is assumed to be constant throughout the flow.

From the expression of \eqref{comi} and \eqref{mqs}, by similar procedure as previous, we get an expression of $\frac{du}{dr}$ as:

\begin{equation}
\label{dudrqs}
\frac{du}{dr}= \frac{u(u^{2}-1)[2r^{3}-2\lambda^{2}(r-2)^{2}+c_{s}^{2}(2r^{3}-2\lambda^{2}r+4\lambda^{2})(3-2r)]}{r(2-r)(u^{2}-c_{s}^{2})(2\lambda^{2}r-4\lambda^{2}-2r^{3})}=\frac{\mathcal{N}}{\mathcal{D}}
\end{equation} 

Here, we will use the same logic as previous that, $u$ being a smooth function of $r$, denominator $\mathcal{D}=0$ implies numerator $\mathcal{N}=0$, which in turn provides the condition for critical points. 

In this case, $\mathcal{D}=0$ indicates that the critical points must coincide with the sonic points, i.e.

\begin{equation}
\label{uqs}
u\mid_{r_{c}} = c_{s}
\end{equation}

Similarly, from $\mathcal{N}=0$, the full critical condition is obtained as,

\begin{equation}
\label{crqs}
u_{c}=c_{s}\mid_{r_{c}}=\sqrt{\frac{2\lambda^{2}(r-2)^{2}-2r^{3}}{(2r^{3}-2\lambda^{2}r+4\lambda^{2})(3-2r)}}
\end{equation}

Simplifying \eqref{crqs} we get an equation in $r_c$ in the following $4th$--order polynomial:

\begin{equation}
\label{polyqs}
f\left(r_c\right)=4 r_c^4 c_s^2-2 r_c^3 \left(3 c_s^2+1\right)-2 \lambda ^2 r_c^2 \left(2 c_s^2-1\right)+2 \lambda ^2 r_c \left(7 c_s^2-4\right)-4 \lambda ^2 \left(3 c_s^2-2\right).
\end{equation}

\begin{figure}
	\begin{center}
		\includegraphics[scale=0.9]{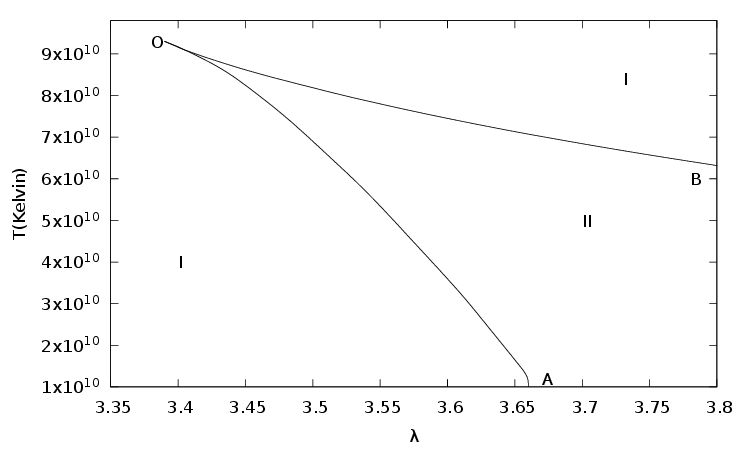}
		\caption{Isothermal accretion in conical disc and related bifurcation phenomena}
		\label{quiso}
	\end{center}
\end{figure}

\noindent
Performing the Sturm analysis as in previous sections, we obtain a bifurcation curve in $\lambda$ - $T$ for the specified range of $\lambda$ and $T$. This corresponding bifurcation diagram is depicted in figure \ref{quiso}.

\noindent
In figure \ref{quiso}, region I denotes the mono-critical region where only one critical point exists and region II denotes the multi-critical region where three critical points exist. Curve OA line represents the left boundary (between region I and II) while OB represents the right boundary (between region II and I). Region OAB (bounded by OA and OB) specifies the multi-critical region II. For instance, at $T = 7.0 \times 10^{10}$ K, in the range of $0\leq\lambda\leq3.50$ (region I) there is a mono-critical solution of $f(r_{c})$ and in the range $3.50\leq\lambda\leq3.67$ there is a multi-critical solution of $f(r_{c})$ (region II) reverting back to mono-critical (mono-transonic) solutions in the range $\lambda\geq3.67$ (region I).

\subsection{Flow in vertical hydrostatic equilibrium}

\subsubsection{Abramowicz-Lanza-Percival accretion model}

The mass accretion rate for this accretion disc geometry is given by,

\begin{equation}
\label{mab1}
 \dot{M}=2\pi\rho r \frac{u \sqrt{1-\frac{2}{r}}}{\sqrt{1-u^{2}}} 2H(r)^{\text{iso}}
\end{equation}
\noindent
where $H(r)^{\text{iso}}$ denotes the half--thickness of the accretion disc.
\noindent
The expression for $H(r)^{\text{iso}}$ is given by, 
\begin{equation}
\label{hra}
H(r)^{\text{iso}}=r c_{s} \frac{\sqrt{2(1-u^{2})(r^{3}-\lambda^{2}(r-2))}}{\lambda\sqrt{r-2}}
\end{equation}
\noindent
Substituting the value of $H(r)^{\text{iso}}$ from equation \eqref{hra} into equation \eqref{mab1} we obtain the final expression of mass accretion rate $\dot{M}$ as,

\begin{equation}
\label{mab}
 \dot{M}=4\pi\rho r^{\frac{3}{2}} \frac{u c_{s}\sqrt{2(r^{3}-\lambda^{2}(r-2))}}{\lambda}  
 \end{equation}
\noindent
Using \eqref{comi} and \eqref{mab}, we obtain,

\begin{equation}
\label{dudra}
\frac{du}{dr}= \frac{u(u^{2}-1)[r^{3}-\lambda^{2}(r-2)^{2}+c_{s}^{2}(2-r)(3r^{3}-2\lambda^{2}r+3\lambda^{2}]}{r(r-2)(r^{3}-\lambda^{2}r+2\lambda^{2})[u^{2}(c_{s}^{2}+1)-c_{s}^{2}]}=\frac{\mathcal{N}}{\mathcal{D}}
\end{equation} 

As $u$ is a continuous function of $r$, by the above mentioned argument, $\mathcal{D}=0$ implies numerator $\mathcal{N}$ must be zero.
\noindent
Here, $\mathcal{D}=0$ gives us,

\begin{equation}
\label{ua}
u\mid_{r_{c}} =\frac {c_{s}}{\sqrt{1+c_{s}^{2}}}
\end{equation}
 
 And the other critical point condition is obtained from $\mathcal{N}=0$ as,
 
 \begin{equation}
 \label{cab}
 c_{s}\mid_{r_{c}}=\sqrt{\frac{\lambda^{2}(r-2)^{2}-r^{3}}{(2-r)(3r^{3}-2\lambda^{2}r+3\lambda^{2})}}
 \end{equation}
 
 By putting \eqref{ua} into \eqref{cab} and simplifying the whole expression, we get a 4th order polynomial in $r_{c}$ as,

\begin{equation}
\label{polyab}
f\left(r_c\right)=6 r_c^4 c_s^2-2r_c^3 \left(6 c_s^2+1\right)+2r_c^2 \lambda ^2 \left(1-2 c_s^2\right)+2r_c \lambda ^2 \left(7 c_s^2-4\right)+4\lambda ^2\left(2 -3 c_s^2\right)
\end{equation}

\begin{figure}
	\begin{center}
		\includegraphics[scale=0.8]{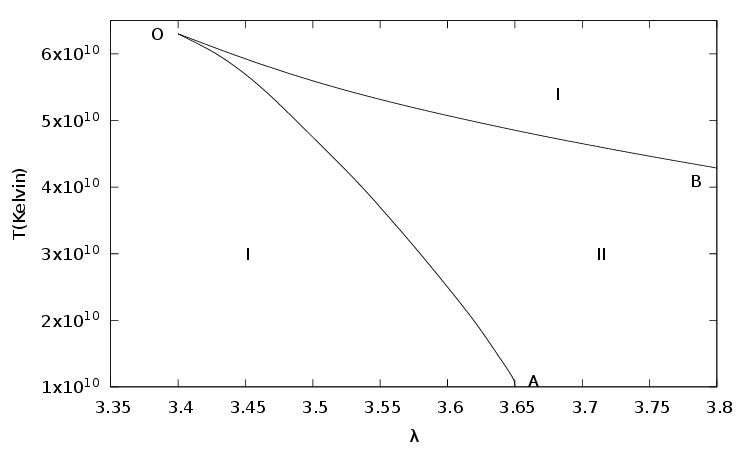}
		\caption{Isothermal accretion in ALP flow model and related bifurcation phenomena}
		\label{apiso}
	\end{center}
\end{figure}

We know that, the number of critical points is equal to the number of real roots (residing outside the horizon) of the above polynomial. Applying Sturm analysis, we generate a bifurcation diagram in $\lambda$ - $T$ space to describe the flow qualitatively. The respective diagram is shown in figure \ref{apiso}. 

In figure \ref{apiso}, the region I denotes the mono-critical region where only one critical point exists and region II denotes the multi-critical region where three critical points exist. Curve OA represents the left boundary (between region I and II) and OB represents the right boundary (between region II and I). Region OAB (bounded by OA and OB) specifies the multi-critical region II. For instance, at $T = 5.0 \times 10^{10}$ K, in the range of $0\leq\lambda\leq3.49$ (region I) there is a mono-critical solution of $f(r_{c})$ and in the range $3.49\leq\lambda\leq3.62$ there is a multi-critical solution of $f(r_{c})$(region II) and in the range $\lambda\geq3.62$ there again exists a mono-critical solution of $f(r_{c})$ (region I).

\subsubsection{Novikov-Thorne accretion flow model}

Here the mass accretion rate is given by,

\begin{equation}
\label{mn1}
\dot{M}=4\pi\rho r \frac{u \sqrt{1-\frac{2}{r}}}{\sqrt{1-u^{2}}} H(r)
\end{equation}
\noindent
where the disc height $H(r)$ is expressed as

\begin{equation}
\label{hrn}
H(r)=c_{s} r^{\frac{3}{2}}\frac{\sqrt{1-\frac{3}{r}}}{\sqrt{1-\frac{2}{r}}}.
\end{equation}
\noindent
Substituting \eqref{hrn} into \eqref{mn1}, the mass accretion rate is obtained as,

\begin{equation}
\label{mn}
\dot{M}=4\pi\rho c_{s} r^{\frac{5}{2}}\frac{u\sqrt{1-\frac{3}{r}}}{\sqrt{1-u^{2}}}
\end{equation}
\noindent
which remains constant throughout the flow.

Using \eqref{comi} and \eqref{mn} we obtain,

\begin{equation}
\nonumber
\frac{du}{dr}= \frac{u(1-u^{2}) [2r^{3}(r-3)-2\lambda^{2}(r^{3}-7r^{2}+16r-12)-c_{s}^{2} r^{3}(5r-12)(r-2)]}{2r(2-r)(r-3)(r^{3}-\lambda^{2}r+2\lambda^{2})(u^{2}-c_{s}^{2})}
\end{equation}

\begin{equation}
\label{dudrn}
+\frac{u(1-u^{2})[\lambda^{2} c_{s}^{2}(5r^{3}-32r^{2}+68r-48)]}{2r(2-r)(r-3)(r^{3}-\lambda^{2}r+2\lambda^{2})(u^{2}-c_{s}^{2})}=\frac{\mathcal{N}}{\mathcal{D}}
\end{equation}

Again using the aforementioned argument that $u$ being a smooth function of $r$, null value of denominator $\mathcal{D}$ would imply that numerator $\mathcal{N}$ should go to zero, we arrive at the criticality conditions.

Inspecting denominator $\mathcal{D}$ we can say that the critical points coincide with the sonic points in the isothermal NT model:

\begin{equation}
\label{urn}
u\mid_{r_{c}} = c_{s}
\end{equation}

$\mathcal{N}=0$ completes the critical condition as:

\begin{equation}
\label{crn}
u_{c}=c_{s}\mid_{r_{c}}=\sqrt{\frac{2r^{3}(r-3)-2\lambda^{2}(r^{3}-7r^{2}+16r-12)}{r^{3}(5r-12)(r-2)-\lambda^{2} (5r^{3}-32r^{2}+68r-48)}}
\end{equation}
\noindent
Now simplifying the above expression \eqref{crn} we get a $5th$--order polynomial in $r_{c}$ as:

\begin{equation}
\nonumber
f\left(r_c\right)=r_c^3 \left(-5 \lambda ^2 c_s^2+24 c_s^2+2 \lambda ^2+6\right)+r_c^2 \left(32 \lambda ^2 c_s^2-14 \lambda ^2\right)
\end{equation}
\begin{equation}
\label{polyn}
+r_c \left(32 \lambda ^2-68 \lambda ^2 c_s^2\right)+5 r_c^5 c_s^2-r_c^4 \left(22 c_s^2+2\right)+\left(48 \lambda ^2 c_s^2-24 \lambda ^2\right)
\end{equation}

\begin{figure}
	\begin{center}
		\includegraphics[scale=0.9]{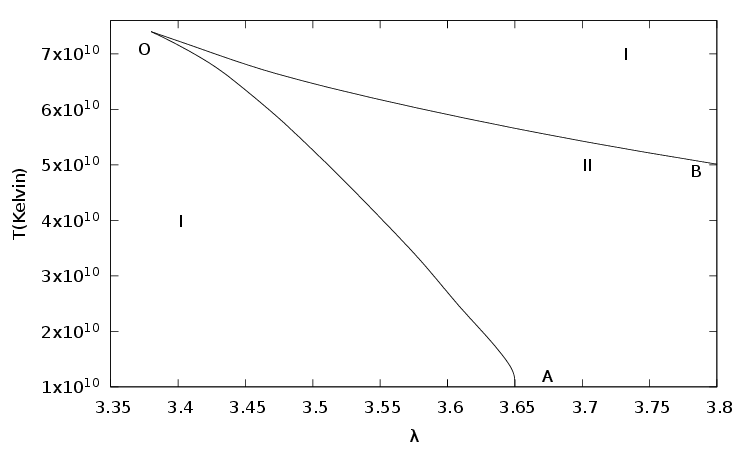}
		\caption{Isothermal accretion in NT flow model and related bifurcation phenomena}
		\label{ntiso}
	\end{center}
\end{figure}

The number of critical points equals to the number of real roots of the above polynomial. Applying Sturm analysis, we generate a bifurcation diagram in $\lambda$ - $T$ space to describe the accretion flow qualitatively. The respective diagram is shown in figure \ref{ntiso}.

In figure \ref{ntiso}, the region I denotes the mono-critical region where only one critical point exists and region II denotes the multi-critical region where three critical points exist. Here, OA represents the left boundary (between region I and II) and OB represents the right boundary (between region II and I). Region OAB (bounded by OA and OB) specifies the multi-critical region II. For instance, at $T = 5.5 \times 10^{10}$ K, in the range of $0\leq\lambda\leq3.49$ (region I) there are mono-critical solutions of $f(r_{c})$ and in the range $3.49\leq\lambda\leq3.69$ there exist multi-critical solutions of $f(r_{c})$ (region II) reverting back to mono-critical physical solutions in the range $\lambda\geq3.69$ (region I).

\subsubsection{Riffert--Herold accretion flow model}

The mass accretion rate for this RH accretion model is:
\begin{equation}
\label{mrh1}
\dot{M}=4\pi\rho r \frac{u \sqrt{1-\frac{2}{r}}}{\sqrt{1-u^{2}}} H(r)
\end{equation}
\noindent
where disc height $H(r)$ is represented by the following expression,
\begin{equation}
\label{hrr}
H(r)=2 r^{\frac{3}{2}} c_{s} \sqrt{1-\frac{3}{r}}
\end{equation}
\noindent
Substituting \eqref{hrr} into \eqref{mrh1}, we obtain,
\begin{equation}
\label{mrh}
\dot{M}=8\pi\rho c_{s}r^{\frac{5}{2}}\frac{u \sqrt{(1-\frac{2}{r})(1-\frac{3}{r}}}{\sqrt{1-u^{2}}}
\end{equation}
\noindent
which remains constant throughout the flow.

\noindent
From the expression of $\dot{M}$ (eq. \eqref{mrh}) and $\xi$ (eq. \eqref{comi}), by using some basic algebra as previous, we get an expression of $\frac{du}{dr}$ which is following:

\begin{equation}
\nonumber
\frac{du}{dr}=\frac{u \left(u^2-1\right) (-5 r^5 c_s^2+20 r^4 c_s^2+5 \lambda ^2 r^3 c_s^2-18 r^3 c_s^2-30 \lambda ^2 r^2 c_s^2+58 \lambda ^2 r c_s^2-36 \lambda ^2 c_s^2)}{r (r-3) (r-2) \left(2 \lambda ^2+r^3-\lambda ^2 r\right) \left(u^2-c_s^2\right)} 
\end{equation}
\begin{equation}
\label{dudrrh}
+\frac{u(u^{2}-1)(24 \lambda ^2+2 r^4-2 \lambda ^2 r^3-6 r^3+14 \lambda ^2 r^2-32 \lambda ^2 r)}{r (r-3) (r-2) \left(2 \lambda ^2+r^3-\lambda ^2 r\right) \left(u^2-c_s^2\right)}=\frac{\mathcal{N}}{\mathcal{D}}
\end{equation}

Now, as $u$ is a smooth function of $r$, following the previously argument, zero value of denominator $\mathcal{D}$ at some radial points in the prescribed range implies that, numerator $\mathcal{N}$ has to be zero at those points, which gives us the condition for criticality in the process of accretion.

Equating $\mathcal{D}$ to zero, we conclude that, for isothermal RH flow, critical points are identical to the sonic points:

\begin{equation}
\label{ucrr}
u\mid_{r_{c}} = c_{s}
\end{equation}

and, then from  $\mathcal{N}=0$, we get the required criticality condition as:

\begin{equation}
\label{crrh}
u_{c}=c_{s}\mid_{r_{c}}=\sqrt{\frac{(24 \lambda ^2+2 r^4-2 \lambda ^2 r^3-6 r^3+14 \lambda ^2 r^2-32 \lambda ^2 r)}{(5 r^5 -20 r^4 -5 \lambda ^2 r^3+18 r^3+30 \lambda ^2 r^2 -58 \lambda ^2 r+36 \lambda ^2)}}
\end{equation}

Simplifying above equation \eqref{crrh} we get a $5th$--order polynomial in $r_{c}$,

\begin{equation}
\nonumber
f\left(r_c\right)\text{=}r_c^3 \left(-5 \lambda ^2 c_s^2+18 c_s^2+2 \lambda ^2+6\right)+r_c^2 \left(30 \lambda ^2 c_s^2-14 \lambda ^2\right)
\end{equation}
\begin{equation}
\label{polyrh}
+r_c \left(32 \lambda ^2-58 \lambda ^2 c_s^2\right)+5 r_c^5 c_s^2-r_c^4 \left(20 c_s^2+2\right)+\left(36 \lambda ^2 c_s^2-24 \lambda ^2\right)
\end{equation}

\begin{figure}
	\begin{center}
		\includegraphics[scale=0.8]{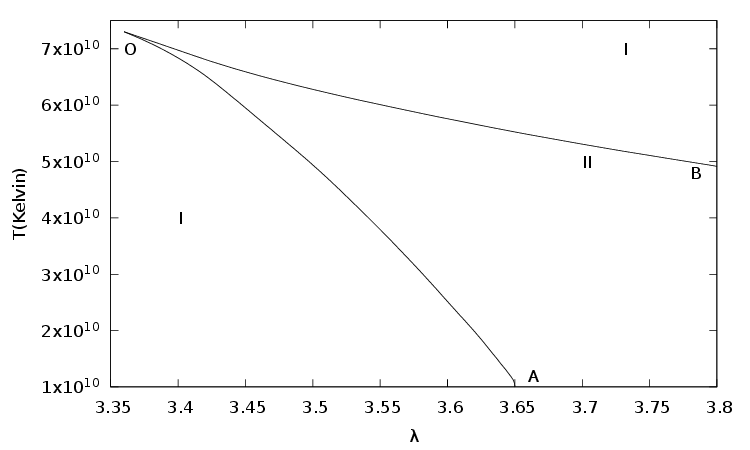}
		\caption{Isothermal accretion in RH flow model and related bifurcation phenomena}
		\label{rhiso}
	\end{center}
\end{figure}

The number of critical points is equal to the number of real roots of the above polynomial. Applying Sturm analysis as earlier, we obtain a bifurcation diagram in $\lambda$ - $T$ space to describe the accretion flow qualitatively (figure \ref{rhiso}). 

In figure \ref{rhiso}, region I denotes the mono-critical region where only one critical point exists and region II denotes the multi-critical region where three critical points exist. Curve OA depicts the left boundary (between region I and II) and OB represents the right boundary (between region II and I). Region OAB (bounded by OA and OB) signifies the multi-critical region II. For instance, at $T = 5.5 \times 10^{10}$ K, in the range of $0\leq\lambda\leq3.48$ (region I) mono-critical solutions of $f(r_{c})$ exist and in the range $3.48\leq\lambda\leq3.66$ there are multi-critical solutions of $f(r_{c})$ (region II) reverting back to mono-critical solutions in the range $\lambda\geq3.66$ (region I). The multi-critical isothermal flow parameter ranges for various disc geometries are provided in table 2.
\begin{table}
\begin{center}
\begin{tabular}{|c| c| c| c|}
\hline \hline
Disc geometry  & $\lambda$ &$\mathcal{T}$ ~~ \\ \hline

CH & 3.67 - 3.80 & $2.0\times 10^{10}-1.8\times 10^{11} $ \\ \hline

Quasi-spherical & 3.66 - 3.80 & $10^{10}-9.3\times 10^{10}$ \\ \hline

ALP & 3.65 - 3.80 & $10^{10}-6.40\times10^{10}$   \\   \hline

NT & 3.65 - 3.80 & $10^{10}-7.4\times10^{10}$   \\   \hline

RH & 3.65 - 3.80 & $10^{10}-7.3\times10^{10}$ \\ \hline
\end{tabular}
\caption{\centering Multi-critical bifurcation regions in isothermal accretion for difference disc geometries}
\end{center}
\end{table}

\section{Concluding remarks}\label{conc}

Analytical investigation of the multi-transonicity for accreting black hole systems, as presented in our work, is based on the fact that one can express the integral solution of the Euler equation as a polynomial of the critical points with suitable constant co-efficients that are functions of initial boundary conditions specifying the flow. If ${\cal E}$ (for polytropic accretion) or $\xi$ (for isothermal accretion) cannot be written as polynomials in $r_c$, Sturm chain cannot be constructed. For accretion onto a rotating black hole, neither ${\cal E}$ nor $\xi$  can be cast into a polynomial in $r_c$ for {\it any} kind of flow geometry governed by any thermodynamic equation of state. Our work, thus, cannot be directly extended to investigate the multi-transonic behaviour in the Kerr metric using full analytical formalism.

Using the methodology as described in the present work, we can at most predict, for which initial boundary conditions the accretion will have three critical points. We, however, cannot predict the nature of the critical points completely analytically since such prediction requires (see, e.g., \cite{gkrd07mnras} for further details) knowledge about the exact location of the critical point, which cannot be determined without the application of numerical computation. However, since we know that no two successive critical points can be of the same type, it is thus evident that out of three critical points, one can either have two saddle types and one centre type or one centre type and two saddle types. A globally valid stationary transonic accretion solution joins infinity with the event horizon. Hence no physical accretion solution can pass through a centre type sonic point. This indicates that a multi-critical solution should have two saddle points at its two extremities and a centre type critical point will be flanked by such saddle points. Thus, although we cannot directly establish the nature of the critical points but we can have a qualitative understanding that would enable us to predict how many critical points of a multi-critical solution will be of what nature.

Analytical solution of critical points exist, however, in certain special cases where the resultant polynomial equation in critical point $r_c$ has an order which is less than or equal to $4$. Such solutions make it possible to obtain numerical values of $r_c$ without the use of computational root-finding algorithms. Numerical values of $r_c$ allow us to readily calculate the values of entropy accretion rate $\dot{\mathcal{M}}$ (polytropic) and quasi-specific energy $\xi$ (isothermal). These two quantities are peculiar in the sense that they remain constant unless encountered by a shock. A shock in the flow would cause a discrete jump in $\dot{\mathcal{M}}$ and $\xi$ to higher values which would again remain unperturbed unless there is a subsequent shock. We know that the outer and inner critical points (the two saddle types) are mediated by a shock transition. Thus, it is evident that if the values of $\dot{\mathcal{M}}$ and $\xi$ are compared at the inner and outer critical points, say $r_c^{in}$ and $r_c^{out}$ respectively, then for an {\it incoming flow} (accretion), $\dot{\mathcal{M}}(r_c^{in})>\dot{\mathcal{M}}(r_c^{out})$ (polytropic) and $\xi(r_c^{in})>\xi(r_c^{out})$ (isothermal). Whereas, for an {\it outgoing flow} (wind), $\dot{\mathcal{M}}(r_c^{in})<\dot{\mathcal{M}}(r_c^{out})$ (polytropic) and $\xi(r_c^{in})<\xi(r_c^{out})$ (isothermal). Consequently, once the location of critical point is known, it is not only possible to determine its nature quantitatively (using analytical formalism which is beyond the scope of our present work), but also to distinguish between the multi-transonic accretion and multi-transonic wind solutions. It may be noted that the polynomial eqns.\ref{polych}, \ref{polyqs} and \ref{polyab} derived in sections $6.1$, $6.2$ and $6.3.1$ are of the $4^{th}$ order. Using the theory of higher algebra, analytical solution to polynomial equations till $4^{th}$ order can be obtained.

It is interesting to note that the qualitative nature of the multi-critical behaviour remains roughly the same for all kinds of geometric configurations of the flow and for all types of thermodynamic equations of state governing the flow. We have also investigated such behaviour (results not presented here) for accretion onto non-rotating black holes under the influence of post-Newtonian pseudo-Schwarzschild black hole potentials (see, e.g., \cite{das02apj} for detailed description of various pseudo-Schwarzschild potentials), and found that the overall qualitative pattern of the transition from mono to multi-transonicity remains roughly the same. It seems, as if, something `conspires' to have identical (in a qualitative sense) flow patterns for accretion onto non-rotating black holes, irrespective of the spacetime (general relativistic or pseudo-Newtonian), disc geometry, or the thermodynamic flow profile. We anticipate that the structure of the governing equations of the autonomous dynamical systems corresponding to the space gradient of the flow velocity and the sound speed has something inherently common to it which gives rise to such striking similarity in the transonic behaviour for accreting black hole systems. Further discussion about this non-trivial issue is beyond the scope of this paper and will be presented elsewhere. 

\backmatter











\begin{appendices}

\section{Sturm Theorem}\label{secA1}

The number of real roots of an algebraic polynomial with real coefficients whose roots are simple over an interval, the endpoints of which are not roots, is equal to the difference between the number of sign changes of the Sturm chains formed for the interval ends. \\

\noindent
Sturm chain -- construction: \\
\noindent
$p$ -- a polynomial with real coefficients,
$P_{i}$ -- i$^{th}$ element of the Sturm chain,
$$P_{0} = p$$
$$P_{1} = p'$$
and for $i\geq 2$
$$P_{i}=−\text{remainder}\left(\frac{p_{i-2}}{p_{i-1}}\right)$$
The number of real roots in a half-open interval (a, b] of the polynomial $\mathcal{P}$ :~~$ \#(\text{change in sign of}$
$P_{i}$ s at a)$ - \#$(change in sign of $P_{i}$ s at b).

\section{Polynomial coefficients}\label{secA2}
\subsection{Polytropic constant height disc}\label{secA2.1}
The polynomial equation for the polytropic accretion with constant height disc structure is given by,
\begin{equation}
\Sigma_{i=0}^{11} a_{ch_{i}}{r_{c}}^{i}=0
\end{equation}
where, the respective coefficients in terms of initial boundary conditions can be expressed as, \\
$a_{ch_{0}}=-36 \mathcal{E}^2 \lambda^6 + 24\mathcal{E}^2 \gamma \lambda^6 - 4\mathcal{E}^2\gamma^2 \lambda^6$, \\
$a_{ch_{1}}=84 \mathcal{E}^2 \lambda^6 -64 \mathcal{E}^2 \gamma\lambda^6 + 12 \mathcal{E}^2 \gamma^2 \lambda^6$ \\
$a_{ch_{2}}=4 \lambda^4 - 8\gamma \lambda^4 + 4 \gamma^2 \lambda^4 - 73 \mathcal{E}^2 \lambda^6 + 62 \mathcal{E}^2  \gamma \lambda^6 - 13\mathcal{E}^2 \gamma^2 \lambda^6$, \\
$a_{ch_{3}}=-16 \lambda^4 + 36 \mathcal{E}^2 \lambda^4 + 32\gamma \lambda^4 -12 \mathcal{E}^2 \gamma \lambda^4 - 16 \gamma^2 \lambda^4 + 28\mathcal{E}^2 \lambda^6 - 26 \mathcal{E}^2 \gamma \lambda^6 \\
+ 6 \mathcal{E}^2  \gamma^2 \lambda^6$, \\
$a_{ch_{4}} =25 \lambda^4 - 72\mathcal{E}^2 \lambda^4 - 50\gamma \lambda^4 +  34  \mathcal{E}^2 \gamma \lambda^4 + 25\gamma^2 \lambda^4 -  2\mathcal{E}^2 \gamma^2 \lambda^4 - 4\mathcal{E}^2 \lambda^6 + 4 \mathcal{E}^2  \gamma \lambda ^6 -\mathcal{E}^2 \gamma^2 \lambda^6$, \\
$a_{ch_{5}}=4\lambda^2 -8 \gamma \lambda^2 + 4\gamma^2 \lambda^2 -19 \lambda^4 +59 \mathcal{E}^2 \lambda^4 + 38\gamma\lambda^4 - 38 \mathcal{E}^2 \gamma \lambda^4 -19 \gamma^2 \lambda^4 + 5 \mathcal{E}^2 \gamma^2 \lambda^4$, \\
$a_{ch_{6}}=-14 \lambda^2 + 28\gamma \lambda^2 -12 \mathcal{E}^2 \gamma \lambda^2 -14 \gamma^2 \lambda^2 + 3\mathcal{E}^2 \gamma^2 \lambda^2 + 7\lambda^4 -\mathcal{E}^2 \lambda^4 - 14 \gamma  \lambda^4 + 20 \mathcal{E}^2 \gamma \lambda^4 + 7 \gamma^2 \lambda^4 -4\mathcal{E}^2 \gamma^2 \lambda^4$, \\
$a_{ch_{7}}= 18 \lambda^2 - 12 \mathcal{E}^2 \lambda^2 - 36 \gamma\lambda^2 +  28 \mathcal{E}^2 \gamma \lambda^2 + 18\gamma^2 \lambda^2 -  8\mathcal{E}^2 \gamma^2 \lambda^2 -\lambda^4 + 4\mathcal{E}^2 \lambda^4 + 2 \gamma \lambda^4 -  4\mathcal{E}^2 \gamma \lambda^4 \gamma^2 \lambda^4 +\mathcal{E}^2 \gamma^2 \lambda^4$, \\
$a_{ch_{8}}=1 - 2\gamma +\gamma^2 - 10 \lambda^2 + 13\mathcal{E}^2 \lambda^2 + 20\gamma \lambda^2 - 22\mathcal{E}^2 \gamma \lambda^2 -10  \gamma^2 \lambda^2 + 7\mathcal{E}^2 \gamma^2 \lambda^2$, \\
$a_{ch_{9}}=-3 + 6 \gamma - 3 \gamma^2 +\mathcal{E}^2 \gamma^2 + 2\lambda^2 - 4\mathcal{E}^2 \lambda^2 - 4 \gamma\lambda^2 +
6\mathcal{E}^2 \gamma \lambda^2 + 2\gamma^2 \lambda^2 - 2\mathcal{E}^2 \gamma^2 \lambda^2$, \\
$a_{ch_{10}}=3 - 6 \gamma + 2 \mathcal{E}^2 \gamma + 3 \gamma^2 - 2  \mathcal{E}^2 \gamma^2$, \\
$a_{ch_{11}}=-1 + \mathcal{E}^2 + 2 \gamma - 2 \mathcal{E}^2 \gamma -\gamma^2 + \mathcal{E}^2  \gamma^2$.

\subsection{Polytropic quasi-spherical disc}\label{secA2.2}
The polynomial equation for the polytropic accretion with conical disc structure is given by,
\begin{equation}
\Sigma_{i=0}^{11} a_{qu_{i}}x^{i}=0
\end{equation}
where, \\
$a_{qu_{0}}=100  \epsilon^2 \lambda^6 -120  \epsilon^2 \gamma \lambda^6 +  36  \epsilon^2 \gamma^2 \lambda^6$, \\
$a_{qu_{1}}=-320 \epsilon^2 \lambda^6 + 392 \epsilon^2 \gamma \lambda^6 - 120 \epsilon^2 \gamma^2 \lambda^6$, \\
$a_{qu_{2}}= 108 \lambda^4 - 216 \gamma \lambda^4 + 108 \gamma^2 \lambda^4 + 401 \epsilon^2 \lambda^6 - 502 \epsilon^2 \gamma \lambda^6 + 157 \epsilon^2 \gamma^2 \lambda^6$, \\
$a_{qu_{3}}= -324 \lambda^4 + 240 \epsilon^2 \lambda^4 + 648 \gamma \lambda^4 - 324 \epsilon^2 \gamma\lambda^4 - 324 \gamma^2 \lambda4 + 108 \epsilon^2 \gamma^2 \lambda]^4 -247 \epsilon^2 \lambda^6 +316 \epsilon^2 \gamma \lambda^6 - 101 \epsilon^2 \gamma^2 \lambda^6$, \\
$a_{qu_{4}}= 387 \lambda^4 - 564 \epsilon^2 \lambda^4 -774 \gamma \lambda^4 + 786 \epsilon^2 \gamma \lambda^4 + 387 \gamma^2 \lambda^4 -270 \epsilon^2 \gamma^2 \lambda^4 +75 \epsilon^2 \lambda^6 - 98 \epsilon^2 \gamma \lambda^6 +32 \epsilon^2 \gamma^2 \lambda^6$, \\
$a_{qu_{5}}=108 \lambda^2 - 216 \gamma \lambda^2 + 108 \gamma^2  \lambda^2 - 230 \lambda^4 + 502 \epsilon^2 \lambda^4 + 460 \gamma \lambda^4 -716 \epsilon^2 \gamma \lambda^4 -230 \gamma^2\lambda^4 + 252 \epsilon^2 \gamma^2 \lambda^4 -9 \epsilon^2 \lambda^6 + 12 \epsilon^2 \gamma \lambda^6 - 4 \epsilon^2 \gamma^2 \lambda^6$, \\
$a_{qu_{6}} =-270 \lambda^2 + 84  \epsilon^2 \lambda^2 + 540 \gamma \lambda^2 -180  \epsilon^2 \gamma \lambda^2 - 270 \gamma^2 \lambda^2 + 81  \epsilon^2 \gamma^2 \lambda^2  + 68 \lambda^4 - 200  \epsilon^2 \lambda^4 - 136 \gamma \lambda^4 +290  \epsilon^2 \gamma \lambda^4 + 68 \gamma^2 \lambda^4 - 104\epsilon^2 \gamma^2 \lambda^4$, \\
$a_{qu_{7}}= (252 \lambda^2 - 180  \epsilon^2 \lambda^2 -504 \gamma \lambda^2 + 348  \epsilon^2 \gamma \lambda^2 + 252 \gamma^2 \lambda^2 -153 \epsilon^2 \gamma^2 \lambda^2 - 8 \lambda^4 + 30\epsilon^2 \lambda^4 + 16 \gamma \lambda^4 - 44 \epsilon^2 \gamma \lambda]^4 -8 \gamma^2 \lambda^4 + 16 \epsilon^2 \gamma^2 \lambda^4)$, \\
$a_{qu_{8}}= 27 - 54  \gamma + 27  \gamma^2 - 104  \lambda^2 +  124  \epsilon^2 \lambda^2 + 208 \gamma \lambda^2 - 224  \epsilon^2 \gamma \lambda^2 - 104 \gamma^2 \lambda^2 + 96  \epsilon^2 \gamma^2 \lambda^2$, \\
$a_{qu_{9}}=-54 + 8  \epsilon^2 + 108 \gamma -  24  \epsilon^2 \gamma - 54 \gamma^2 + 18  \epsilon^2 \gamma^2 + 16 \lambda^2 -  28  \epsilon^2 \lambda^2 - 32 \gamma \lambda^2 +  48  \epsilon^2 \gamma \lambda^2 + 16 \gamma^2 \lambda^2 - 20  \epsilon^2 \gamma^2 \lambda^2$, \\
$a_{qu_{10}}=36 - 16 \epsilon^2 - 72 \gamma + 40 \epsilon^2 \gamma + 36 \gamma^2 - 24 \epsilon^2 \gamma^2$, \\
$a_{qu_{11}}= -8 + 8 \epsilon^2 + 16 \gamma - 16 \epsilon^2 \gamma - 8  \gamma^2 + 8  \epsilon^2 \gamma^2$.

\subsection{Polytropic Novikov-Thorne disc}\label{secA2.3}
The polynomial equation for the polytropic accretion in Novikov-Thorne disc is given by,
\begin{equation}
\Sigma_{i=0}^{15} a_{NT_{i}}x^{i}=0
\end{equation}
where, \\
$a_{NT_{0}}= 86400 \epsilon^2 \lambda^6 -103680 \gamma \epsilon^2 \lambda^6 +31104 \gamma^2 \epsilon^2 \lambda^6$, \\
$a_{NT_{1}}= -371520 \epsilon^2 \lambda^6 + 450432 \gamma \epsilon^2 \lambda^6 -136512 \gamma^2 \epsilon^2 \lambda^6$, \\
$a_{NT_{2}}= 110592 \lambda^4 - 221184 \gamma \lambda^4 + 110592 \gamma^2 \lambda^4 + $$ $$709344 \epsilon^2 \lambda^6 - 869184 \gamma \epsilon^2 \lambda^6 +266208 \gamma^2 \epsilon^2 \lambda^6$, \\
$a_{NT_{3}}=-470016 \lambda^4 + 940032 \gamma \lambda^4 - 470016 \gamma^2 \lambda^4 + 168480 \epsilon^2 \lambda^4 - 243648 \gamma \epsilon^2 \lambda^4 + 85536 \gamma^2 \epsilon^2 \lambda^4 - 789264 \epsilon^2 \lambda^6 + 977760 \gamma \epsilon^2 \lambda^6 -302736 \gamma^2 \epsilon^2 \lambda^6$, \\
$a_{NT_{4}}= 887040 \lambda^4 - 1774080 \gamma \lambda^4 + 887040 \gamma^2 \lambda^4 - 624672 \epsilon^2 \lambda^4 + 907200 \gamma \epsilon^2 \lambda^4 - 320544 \gamma^2 \epsilon^2 \lambda^4 + 563976 \epsilon^2 \lambda^6 -706608 \gamma \epsilon^2 \lambda^6 + 221256 \gamma^2 \epsilon^2 \lambda^6$, \\
$a_{NT_{5}}= 110592 \lambda^2 - 221184 \gamma \lambda^2 +  110592 \gamma^2 \lambda^2 - 975680 \lambda^4 + 1951360 \gamma \lambda^4 - 975680 \gamma^2 \lambda^4 + 1013472 \epsilon^2 \lambda^4-1478016 \gamma \epsilon^2 \lambda^4 + 525600 \gamma^2 \epsilon^2 \lambda^4 - 268380 \epsilon^2 \lambda^6 + 340200 \gamma \epsilon^2 \lambda^6 -107772 \gamma^2 \epsilon^2 \lambda^6$, \\
$a_{NT_{6}} =-414720 \lambda^2 + 829440 \gamma \lambda^2 -414720 \gamma^2 \lambda^2 + 86184 \epsilon^2 \lambda^2 - 169776 \gamma \epsilon^2 \lambda^2 + 75816 \gamma^2 \epsilon^2 \lambda^2 + 689280 \lambda^4 - 1378560 \gamma \lambda^4 + 689280 \gamma^2 \lambda^4 - 939744 \epsilon^2 \lambda^4 + 1376256 \gamma \epsilon^2 \lambda^4 - 492576 \gamma^2 \epsilon^2 \lambda^4 + 85050 \epsilon^2 \lambda^6 - 109116 \gamma \epsilon^2 \lambda^6 + 34986 \gamma^2 \epsilon^2 \lambda^6$, \\
$a_{NT_{7}}=679680 \lambda^2 - 1359360 \gamma \lambda^2 + 679680 \gamma^2 \lambda^2 - 278964 \epsilon^2 \lambda^2 + 535032 \gamma \epsilon^2 \lambda^2 - 236628 \gamma^2 \epsilon^2 \lambda^2 - 324336 \lambda^4 + 648672 \gamma \lambda^4 - 324336 \gamma^2 \lambda^4 + 544698 \epsilon^2 \lambda^4 - 801132 \gamma \epsilon^2 \lambda^4 + 288602 \gamma^2 \epsilon^2 \lambda^4 - 17307 \epsilon^2 \lambda^6 +22482 \gamma \epsilon^2 \lambda^6 -7299 \gamma^2 \epsilon^2 \lambda^6$, \\
$a_{NT_{8}}= 27648 - 55296 \gamma + 27648 \gamma^2 - 635840 \lambda^2 + 1271680 \gamma \lambda^2 - 635840 \gamma^2 \lambda^2 + 386046 \epsilon^2 \lambda^2 - 723204 \gamma \epsilon^2 \lambda^2 + 101648 \lambda^4 - 203296 \gamma \lambda^4 + 101648 \gamma^2 \lambda^4 - 202086 \epsilon^2 \lambda^4 + 298548 \gamma \epsilon^2 \lambda^4 - 108262 \gamma^2 \epsilon^2 \lambda^4 + 2052 \epsilon^2 \lambda^6 + 316926 \gamma^2 \epsilon^2 \lambda^2 - 2700 \gamma \epsilon^2 \lambda^6 + 888 \gamma^2 \epsilon^2 \lambda^6$, \\
$a_{NT_{9}}= -89856 + 179712 \gamma - 89856 \gamma^2 + 13230 \epsilon^2 - 34020 \gamma \epsilon^2 + 21870 \gamma^2 \epsilon^2 + 371360 \lambda^2 - 742720 \gamma \lambda^2 + 371360 \gamma^2 \lambda^2 - 296199 \epsilon^2 \lambda^2 + 543594 \gamma \epsilon^2 \lambda^2 - 236175 \gamma^2 \epsilon^2 \lambda^2 - 20460 \lambda^4 + 40920 \gamma \lambda^4 - 20460 \gamma^2 \lambda^4 + 46863 \epsilon^2 \lambda^4 - 69558 \gamma \epsilon^2 \lambda^4 + 25395 \gamma^2 \epsilon^2 \lambda^4 - 108 \epsilon^2 \lambda^6 + 144 \gamma \epsilon^2 \lambda ^6 - 48 \gamma^2 \epsilon^2 \lambda^6$, \\
$a_{NT_{10}}= 124992 - 249984 \gamma + 124992 \gamma^2 - 37044 \epsilon^2 + 90216 \gamma \epsilon^2 - 54756 \gamma^2 \epsilon^2 - 138656 \lambda^2 + 277312 \gamma \lambda^2 - 138656 \gamma^2 \lambda^2 + 136134 \epsilon^2 \lambda^2 - 245412 \gamma \epsilon^2 \lambda^2 +105782 \gamma^2 \epsilon^2 \lambda^2 + 2400 \lambda^4 -4800 \gamma \lambda^4 + 2400 \gamma^2 \lambda^4 -6210 \epsilon^2 \lambda^4 + 9264 \gamma \epsilon^2 \lambda^4 -3406 \gamma^2 \epsilon^2 \lambda^4$, \\
$a_{NT_{11}}= -96464 + 192928 \gamma - 96464 \gamma^2 +42903 \epsilon^2 - 99450 \gamma \epsilon^2 + 57423 \gamma^2 \epsilon^2 + 32320 \lambda^2 - 64640 \gamma \lambda^2 + 32320 \gamma^2 \lambda^2 - 37491 \epsilon^2 \lambda^2 + 66554 \gamma \epsilon^2 \lambda^2 -28483 \gamma^2 \epsilon^2 \lambda^2 - 125 \lambda^4 + 250 \gamma \lambda^4 - 125 \gamma^2 \lambda^4 + 360 \epsilon^2 \lambda^4 -540 \gamma \epsilon^2 \lambda^4 + 200 \gamma^2 \epsilon^2 \lambda^4$, \\
$a_{NT_{12}}= 44608 - 89216 \gamma + 44608 \gamma^2 -26334 \epsilon^2 + 58404 \gamma \epsilon^2 - 32286 \gamma^2 \epsilon^2 - 4300 \lambda^2 + 8600 \gamma \lambda^2 - 4300 \gamma^2 \lambda^2 + 5730 \epsilon^2 \lambda^2 -10040 \gamma \epsilon^2 \lambda^2 +4270 \gamma^2 \epsilon^2 \lambda^2$, \\
$a_{NT_{13}}= -12360 + 24720 \gamma - 12360 \gamma^2 + 9045 \epsilon^2 - 19290 \gamma \epsilon^2 + 10265 \gamma^2 \epsilon^2 + 250 \lambda^2 - 500 \gamma \lambda^2 + 250 \gamma^2 \lambda^2 - 375 \epsilon^2 \lambda^2 + 650 \gamma \epsilon^2 \lambda^2 - 275 \gamma^2 \epsilon^2 \lambda^2$, \\
$a_{NT_{14}} = 1900 - 3800 \gamma + 1900 \gamma^2 - 1650 \epsilon^2 + 3400 \gamma \epsilon^2 - 1750 \gamma^2 \epsilon^2$, \\
$a_{NT_{15}}= -125 + 250 \gamma - 125 \gamma^2 + 125 \epsilon^2 -250 \gamma \epsilon^2 + 125 \gamma^2 \epsilon^2$.

\subsection{Polytropic Riffert-Herold disc}\label{secA2.4}
The polynomial equation for the polytropic accretion in Riffert-Herold disc is given by,
\begin{equation}
\Sigma_{i=0}^{14} a_{RH_{i}}x^{i}=0
\end{equation}
where, \\
$a_{RH_{0}}= -3456 \gamma ^2 \epsilon ^2 \lambda ^6+13824 \gamma  \epsilon ^2 \lambda ^6-13824 \epsilon ^2 \lambda ^6$\\
$a_{RH_{1}}= \left(17856 \gamma ^2 \epsilon ^2 \lambda ^6-68544 \gamma  \epsilon ^2 \lambda ^6+65664 \epsilon ^2 \lambda ^6\right)$\\ 
$a_{RH_{2}}= (-39096 \gamma ^2 \epsilon ^2 \lambda ^6+143952 \gamma  \epsilon ^2 \lambda ^6 -132120 \epsilon ^2 \lambda ^6-23328 \gamma ^2 \lambda ^4+46656 \gamma  \lambda ^4-23328 \lambda ^4)$\\
$a_{RH_{3}}=(47472 \gamma ^2 \epsilon ^2 \lambda ^6-167856 \gamma \epsilon ^2 \lambda ^6+147904 \epsilon ^2 \lambda ^6 +101088 \gamma ^2 \lambda ^4
-12960 \gamma ^2 \epsilon ^2 \lambda ^4+44064 \gamma  \epsilon ^2 \lambda ^4-36288 \epsilon ^2 \lambda ^4- 202176 \gamma  \lambda ^4+101088 \lambda ^4$\\
$a_{RH_{4}}= (-35034 \gamma ^2 \epsilon ^2 \lambda ^6+119228 \gamma  \epsilon ^2 \lambda ^6-101146 \epsilon ^2 \lambda ^6-189432 \gamma ^2 \lambda ^4
+53784 \gamma ^2 \epsilon ^2 \lambda ^4-173664 \gamma  \epsilon ^2 \lambda ^4+135432 \epsilon ^2 \lambda ^4+378864 \gamma  \lambda ^4 -189432 \lambda ^4)$\\
$a_{RH_{5}}=(16126 \gamma ^2 \epsilon ^2 \lambda ^6-52964 \gamma  \epsilon ^2 \lambda ^6+43390 \epsilon ^2 \lambda ^6
+200480 \gamma ^2 \lambda ^4-93060 \gamma ^2 \epsilon ^2 \lambda ^4+287496 \gamma  \epsilon ^2 \lambda ^4-215028 \epsilon ^2 \lambda ^4 -400960 \gamma  \lambda ^4+200480 \lambda ^4-23328 \gamma ^2 \lambda ^2+46656 \gamma  \lambda ^2-23328 \lambda ^2$\\
$a_{RH_{6}}= (-4531 \gamma ^2 \epsilon ^2 \lambda ^6+ 14402 \gamma  \epsilon ^2 \lambda ^6-11427 \epsilon ^2 \lambda ^6-131060 \gamma ^2 \lambda ^4+87168 \gamma ^2 \epsilon ^2 \lambda ^4 -259536 \gamma  \epsilon ^2 \lambda ^4+187872 \epsilon ^2 \lambda ^4+262120 \gamma  \lambda ^4-131060 \lambda ^4+89424 \gamma ^2 \lambda ^2 -14742 \gamma ^2 \epsilon ^2 \lambda ^2+36612 \gamma  \epsilon ^2 \lambda ^2-18630 \epsilon ^2 \lambda ^2-178848 \gamma  \lambda ^2
+89424 \lambda ^2)$\\
$a_{RH_{7}}= (712 \gamma ^2 \epsilon ^2 \lambda ^6-2196 \gamma  \epsilon ^2 \lambda ^6+1692 \epsilon ^2 \lambda ^6 +54200 \gamma ^2 \lambda ^4-47838 \gamma ^2 \epsilon ^2 \lambda ^4+138124 \gamma  \epsilon ^2 \lambda ^4 -97398 \epsilon ^2 \lambda ^4-108400 \gamma  \lambda ^4+54200 \lambda ^4-144720 \gamma ^2 \lambda ^2 +48006 \gamma ^2 \epsilon ^2 \lambda ^2-116100 \gamma  \epsilon ^2 \lambda ^2+61398 \epsilon ^2 \lambda ^2 +289440 \gamma  \lambda ^2-144720 \lambda ^2)$\\
$a_{RH_{8}}= (-48 \gamma ^2 \epsilon ^2 \lambda ^6+144 \gamma  \epsilon ^2 \lambda ^6 -108 \epsilon ^2 \lambda ^6-13850 \gamma ^2 \lambda ^4+15415 \gamma ^2 \epsilon ^2 \lambda ^4 -43374 \gamma  \epsilon ^2 \lambda ^4+29931 \epsilon ^2 \lambda ^4+27700 \gamma  \lambda ^4-13850 \lambda ^4 +128120 \gamma ^2 \lambda ^2-63549 \gamma ^2 \epsilon ^2 \lambda ^2++151614 \gamma  \epsilon ^2 \lambda ^2 -82485 \epsilon ^2 \lambda ^2-256240 \gamma  \lambda ^2+128120 \lambda ^2-5832 \gamma ^2+11664 \gamma -5832)$\\
$a_{RH_{9}}= (2000 \gamma ^2 \lambda ^4-2706 \gamma ^2 \epsilon ^2 \lambda ^4+7448 \gamma  \epsilon ^2 \lambda ^4 -5046 \epsilon ^2 \lambda ^4-4000 \gamma  \lambda ^4+2000 \lambda ^4-67000 \gamma ^2 \lambda ^2 +43914 \gamma ^2 \epsilon ^2 \lambda ^2-104124 \gamma  \epsilon ^2 \lambda ^2+57882 \epsilon ^2 \lambda ^2 +134000 \gamma  \lambda ^2-67000 \lambda ^2+19440 \gamma ^2-5292 \gamma ^2 \epsilon ^2+7560 \gamma  \epsilon ^2 -2700 \epsilon ^2-38880 \gamma +19440$\\ 
$a_{RH_{10}}= (-125 \gamma ^2 \lambda ^4+ 200 \gamma ^2 \epsilon ^2 \lambda ^4-540 \gamma  \epsilon ^2 \lambda ^4 +360 \epsilon ^2 \lambda ^4+250 \gamma  \lambda ^4-125 \lambda ^4+20700 \gamma ^2 \lambda ^2-16749 \gamma ^2 \epsilon ^2 \lambda ^2+ 39622 \gamma  \epsilon ^2 \lambda ^2-22389 \epsilon ^2 \lambda ^2-41400 \gamma  \lambda ^2+20700 \lambda ^2 -26460 \gamma ^2+12789 \gamma ^2 \epsilon ^2-20286 \gamma  \epsilon ^2+7965 \epsilon ^2+52920 \gamma -26460$\\
$a_{RH_{11}}= (-12222 \epsilon ^2 \gamma ^2+3350 \epsilon ^2 \lambda ^2 \gamma ^2-3500 \lambda ^2 \gamma ^2+18800 \gamma ^2+21036 \epsilon ^2 \gamma -7920 \epsilon ^2 \lambda ^2 \gamma +7000 \lambda ^2 \gamma -37600 \gamma -8982 \epsilon ^2+4530 \epsilon ^2 \lambda ^2-3500 \lambda ^2+18800)$\\
$a_{RH_{12}}= 5775 \epsilon ^2 \gamma ^2-275 \epsilon ^2 \lambda ^2 \gamma ^2+250 \lambda ^2 \gamma ^2-7350 \gamma ^2-10590 \epsilon ^2 \gamma +650 \epsilon ^2 \lambda ^2 \gamma -500 \lambda ^2 \gamma +14700 \gamma +4835 \epsilon ^2-375 \epsilon ^2 \lambda ^2+250 \lambda ^2-7350$\\
$a_{RH_{13}}= (-1350 \epsilon ^2 \gamma ^2+1500 \gamma ^2+2600 \epsilon ^2 \gamma -3000 \gamma -1250 \epsilon ^2+1500)$\\
$a_{RH_{14}}= (125 \epsilon ^2 \gamma ^2-125 \gamma ^2-250 \epsilon ^2 \gamma +250 \gamma +125 \epsilon ^2-125)$\\




\end{appendices}



\begin{thebibliography}{99}

\bibitem{fukue83pasj} Fukue, J. PASJ {\bf{35}} (3), 355 (1983).
\bibitem{lu85aa} Lu, J. F. A\& A, {\bf{148}}, 176 (1985).
\bibitem{lu86grg} Lu, J. F. General Rel. \& Grav. {\bf{18}} (1), 45 (1986).
\bibitem{fukue87pasj} Fukue, J. PASJ {\bf{39}} (2), 309 (1987).
\bibitem{ky94mnras} Kafatos, M., Yang, R. X. MNRAS {\bf{268}}, 925 (1994).
\bibitem{nakayama94mnras} Nakayama, K. MNRAS {\bf{270}}, 871 (1994).
\bibitem{yk95aa} Yang, R., Kafatos, M.: A\& A {\bf{295}}, 238 (1995).
\bibitem{chakrabarti96mnras} Chakrabarti, S. K. MNRAS {\bf{283}}, 325 (1996).
\bibitem{pariev96mnras} Pariev, V. I. MNRAS {\bf{283}} (4), 1264 (1996).
\bibitem{lyyy97aa} Lu, J. F., Yu, K. N., Yuan, F. \& Young, E. C. M. A\& A {\bf{321}}, 665 (1997).
\bibitem{pa97mnras} Peitz, J. \& Appl, S. MNRAS {\bf{286}} (3), 681 (1997).
\bibitem{ct98apj} Caditz, D. M. \& Tsuruta, S. ApJ {\bf{501}} (1), 242 (1998).
\bibitem{cd01mnras} Chakrabarti, S. K. \& Das, S. MNRAS {\bf{327}} (3), 808 (2001).
\bibitem{trft02apj} Takahashi, M., Rilett, D., Fukumara, K. \& Tsuruta, S. ApJ {\bf{572}} (2), 950 (2002).
\bibitem{bdw04apj} Barai, P., Das, T. K. \& Wiita, P. J. ApJL {\bf{613}} (1), L49 (2004).
\bibitem{ny08apj} Nagakura, H. \& Yamada, S. ApJ {\bf{689}} (1), 391 (2008).
\bibitem{ny09apj} Nagakara, H. \& Yamada, S. ApJ {\bf{696}} (2), 2026 (2009).
\bibitem{dc12mnras} Das, T. K. \& Czerny, B. MNRAS Lett. {\bf{421}} (1), L24 (2012).
\bibitem{bcdn14cqg} Bilic, N., Choudhary, A., Das, T. K. \& Nag, S. Classical \& Quantum Grav. {\bf{31}} (3), 035002 (2014).
\bibitem{dnhbmcbwkn15na} Das, T. K., Nag, S., Hegde, S., Bhattacharya, S., Maity, I., Czerny, B., Barai, P., Wiita, P. J., Karas, V. \& Naskar, T. New Astronomy {\bf{37}}, 81 (2015).
\bibitem{td15ijmpd} Tarafdar, P. \& Das, T. K. IJMPD {\bf{24}} (14), 1550096 (2015).
\bibitem{abd15grg} Ananda, D. B., Bhattacharya, S. \& Das, T. K. General Rel. \& Grav. {\bf{47}}, 96 (2015).
\bibitem{sj15mnras} Sukova, P. \& Janiuk, A. MNRAS {\bf{447}} (2), 1565 (2015).
\bibitem{lwwbp16apj} Le, T., Wood, K. S., Wolff, M. T., Becker, P. A. \& Putney, J. ApJ {\bf{819}} (2), 112 (2016).
\bibitem{ssnrd16na} Saha, S., Sen, S., Nag, S., Raychowdhury, S. \& Das, T. K.: New Astronomy {\bf{42}}, 10 (2016).
\bibitem{scj17mnras} Sukova, P., Charzynski, S. \& Janiuk, A. MNRAS {\bf{472}} (4), 4327 (2017).
\bibitem{sfd17cqg} Shaikh, M. A., Firdousi, I. \& Das, T. K.: Classical \& Quantum Grav. {\bf{34}} (15), 155008 (2017).
\bibitem{bbd17na} Bollimpalli, D. A., Bhattacharya, S. \& Das, T. K. New Astronomy {\bf{51}}, 153 (2017).
\bibitem{shaikh18cqg} Shaikh, M. A. Classical \& Quantum Grav. {\bf{35}} (5), 055002 (2018).
\bibitem{td18ijmpd} Tarafdar, P. \& Das, T. K. IJMPD {\bf{27}} (3), 1850023 (2018).
\bibitem{dsd18na} Datta, S., Shaikh, M. A. \& Das, T. K. New Astronomy {\bf{63}}, 65 (2018).
\bibitem{ddmc18prd} Dihingia, I. K., Das, S. Maity, D. \& Chakrabarti, S. Phys. Rev. D {\bf{98}} (8), 083004 (2018).
\bibitem{mnd18mnras} Majumder, S. Nag, S. \& Das, T. K. MNRAS {\bf{480}} (3), 3017 (2018).
\bibitem{sd18jaa} Sarkar, B. \& Das, S. Journal of Astrophysics \& Astronomy {\bf{39}} (1), 3 (2018).
\bibitem{sd18prd} Shaikh, M. A. \& Das, T. K. Phys. Rev. D {\bf{98}} (12), 123022 (2018).
\bibitem{smnd19na} Shaikh, M. A., Maity, S., Nag, S. \& Das, T. K. New Astronomy {\bf{69}}, 48 (2019).
\bibitem{ddmn19mnras} Dihingia, I. K., Das, S., Maity, D. \& Nandi, A. MNRAS {\bf{488}} (2), 2412 (2019).
\bibitem{ddn19mnras} Dihingia, I. K., Das, S. \& Nandi, A. MNRAS {\bf{484}} (3), 3209 (2019).
\bibitem{pjs19mnras} Palit, I., Januk, A. \& Sukova, P. MNRAS {\bf{487}} (1), 755 (2019).
\bibitem{pjs19hepr} Palit, I., Janiuk, A. \& Sukova, P. Proceedings of High Energy Phenomena in Relativistic Outflows VII, {\bf{68}} (2019).
\bibitem{pjc20apj} Palit, I., Janiuk, A. \& Czerny, B. ApJ {\bf{904}} (1), 21 (2020).
\bibitem{dmcd20mnras} Dihingia, I. K., Maity, D., Chakrabarti, S. \& Das, S. MNRAS {\bf{102}} (2), 023012 (2020).
\bibitem{soa21raa} Singh, C. B., Okuda, T. \& Aktar, R. Research in Astronomy \& Astrophysics {\bf{21}} (6), 134 (2021).
\bibitem{tmd21prd} Tarafdar, P., Maity, S. \& Das, T. K. Phys. Rev. D {\bf{103}} (2), 023023 (2021).
\bibitem{rb02pre} Ray, A.K. \& Bhattacharjee, J.K. Phys. Rev. E {\bf{66}}, 066303 (2002).
\bibitem{ray03mnrasa} Ray, A.K. MNRAS {\bf{344}}, 83 (2003).
\bibitem{ray03mnrasb} Ray, A.K. MNRAS {\bf{344}}, 1085 (2003).
\bibitem{rb05arxiv} Ray, A.K. \& Bhattacharjee, J.K. {\it{A dynamical systems approach to a thin accretion disc and its time-dependent behaviour on large length scales}} Eprint arXiv:astro-ph/0511018v1 (2005).
\bibitem{rb05apj} Ray, A.K. \& Bhattacharjee, J.K. ApJ {\bf{627}}, 368 (2005).
\bibitem{rb06ijp} Ray, A.K. \& Bhattacharjee, J.K. Indian J. Phys. {\bf{80}}, 1123 (2006).
\bibitem{rb07cqg} Ray, A.K. \& Bhattacharjee, J.K. Classical \& Quantum Grav. {\bf{24}}, 1479 (2007).
\bibitem{br07apj} Bhattacharjee, J.K. \& Ray, A.K. ApJ {\bf{668}}, 409 (2007).
\bibitem{bbdr09mnras} Bhattacharjee, J.K., Bhattacharya, A., Das, T.K. \& Ray, A.K. MNRAS {\bf{398}}, 841 (2009).
\bibitem{addn12grg} Agarwal, S., Das, T. K., Dey, R. \& Nag, S. General Rel. and Grav. {\bf{44}} (7), 1637 (2012).
\bibitem{td18na} Tarafdar, P. \& Das, T. K.  New Astronomy {\bf{62}}, 1 (2018).
\bibitem{tbnd19prd} Tarafdar, P., Bollimpalli, D. A., Nag, S. \& Das, T. K. Phys. Rev. D {\bf{100}}, 043024 (2019).
\bibitem{is75aa} Illarionov, A.F. \& Sunyaev, R. A. A\& A {\bf{39}}, 205 (1975).
\bibitem{ln84aa} Liang, E. P. T. \& Nolan, P. L. Space. Sci. Rev. {\bf{38}}, 353 (1984).
\bibitem{illarionov88sa} Illarionov, A. F. Soviet Astron. {\bf{31}}, 618 (1988).
\bibitem{bbckm98mnras} Bisikalo, A. A., Boyarchuk, V. M., Chechetkin, V. M., Kuznetsov, O. A. \& Molteni, D. MNRAS {\bf{300}}, 39 (1998).
\bibitem{ho99} Ho, L. C., {\it{Observational Evidence For Black Holes in the Universe}}, ed. Chakrabarti, S. K (Dordrecht: Kluwer), 153 (1999)
\bibitem{ia99mnras} Igumenshchev, I. V. \& Abramowicz, M. A. MNRAS {\bf{303}}, 309 (1999).
\bibitem{ablp} M. A. Abramowicz, A. Lanza, and M. J. Percival. ApJ, 479:179, 1997
\bibitem{bcr91} Bochnak, J., Coste, M. \& Roy, M.F. {\it{Real Algebraic Geometry}} Springer, Berlin (1991).
\bibitem{katsu} Matsumoto, R., Kato, S., Fukue, J., $\&$ Okazaki, A. T.:
 PASJ 36, 1 (1984)
 \bibitem{maa}Abramowicz, M.A., Zurek, W.H.: ApJ. 246, 314 (1981)
\bibitem{blabla}Blaes, O., 1987. MNRAS 227, 975.
 \bibitem{noth} Novikov, I. \& Thorne, K. {\it{Astrophysics of Black Holes, Les Astres Occlus}} 343 (1973)
\bibitem{rih} Riffert, H. \& Herold, H. ApJ {\bf{450}}, 508 (1995)
\bibitem{gkrd07mnras} Goswami, S., Khan, S. N., Ray, A. K. \& Das, T. K. MNRAS {\bf{378}} (4), 1407 (2007).
\bibitem{das02apj} Das, T. K. ApJ {\bf{577}} (2), 880 (2002).

\end{thebibliography}


\end{document}